\documentclass[oneside,11pt]{article}

\usepackage[utf8]{inputenc}
\usepackage[T1]{fontenc}

\usepackage{amsthm}
\usepackage{amsmath}
\usepackage{amsfonts}
\usepackage{amssymb}

\usepackage{graphicx} 
\usepackage{float}
\usepackage{booktabs}
\usepackage{multirow}
\usepackage{rotating}
\usepackage{array}

\usepackage{breakcites}

\usepackage{enumitem}

\usepackage[top=1.5cm,bottom=2cm,right=2.5cm,left=2.5cm]{geometry}
\usepackage{titling}

\usepackage{natbib}

\usepackage{ifplatform}

\ifwindows
\fi

\ifwindows
	\usepackage{bbm}
\else
	\iflinux
		\usepackage{bbm}
	\else
		\usepackage{bbm}
	\fi
\fi

\newcommand{\lp}{\left(}
\newcommand{\rp}{\right)}

\newcommand{\R}{\mathbb{R}}

\newcommand{\bx}{\mathbf{x}}

\newcommand{\by}{\mathbf{y}}
\newcommand{\bX}{\mathbf{X}}

\newcommand{\bmu}{\boldsymbol\mu}

\newcommand{\lrp}[1]{\left(#1\right)}
\newcommand{\lrc}[1]{\left[#1\right]}
\newcommand{\lrb}[1]{\left\{#1\right\}}

\newcommand{\norm}[1]{\left|\left| #1\right|\right|}

\newcommand{\Om}[1]{\Omega_{#1}}
\newcommand{\om}[1]{\omega_{#1}}
\newcommand{\Iq}[2]{\int_{\Omega_{q}} #1\,\omega_{q}(d #2)}
\newcommand{\Iqr}[3]{\int_{\Omega_{q}\times\R} #1\,d #3\,\omega_{q}(d #2)}

\newcommand{\Ir}[2]{\int_{\R} #1\,d #2}

\DeclareFontFamily{OT1}{pzc}{}
\DeclareFontShape{OT1}{pzc}{m}{it}{<-> s * [1.10] pzcmi7t}{}
\DeclareMathAlphabet{\mathpzc}{OT1}{pzc}{m}{it}

\usepackage[pdftex,final,bookmarksnumbered,bookmarksopen=false,breaklinks,colorlinks]{hyperref}
\hypersetup{
	final,
    bookmarksnumbered=true,
    bookmarksopen=true,
    bookmarksopenlevel=0,
    unicode=false,          
    pdftoolbar=true,        
    pdfmenubar=true,        
    pdffitwindow=false,     
    pdftitle={A test for directional-linear independence, with applications to wildfire orientation and size},    
	pdfdisplaydoctitle=true, 
	pdftoolbar=true,
	pdfmenubar=true,
	pdflang={English},
    pdfauthor={Eduardo Garcia-Portugues, Ana M. G. Barros, Rosa M. Crujeiras, Wenceslao Gonzalez-Manteiga, J. M. C. Pereira},     
    pdfsubject={arXiv paper},   
    pdfcreator={Eduardo Garcia-Portugues},   
    pdfproducer={Eduardo Garcia-Portugues}, 
    pdfkeywords={Bootstrap} {Directional-linear density}  {Independence test} {Nonparametric estimation} {Permutations} {Wildfires data}, 
    pdfnewwindow=true,      
    breaklinks=true,
    hidelinks,       
    linkcolor=black,          
    citecolor=black,        
    filecolor=magenta,      
    urlcolor=cyan           
}

\newcommand{\co}{\addtocounter{equation}{1}\arabic{equation}}

\newtheorem{lem}{Lemma}
\newtheorem{algo}{Algorithm}

\allowdisplaybreaks

\begin{document}

\title{A test for directional-linear independence, with applications to wildfire orientation and size}
\setlength{\droptitle}{-1cm}
\predate{}%
\postdate{}%
\author{Eduardo Garc\'ia-Portugu\'es$^{1,3}$, Ana M. G. Barros$^{2}$, Rosa M. Crujeiras$^{1}$,\\ Wenceslao Gonz\'alez-Manteiga$^{1}$, and J. M. C. Pereira$^{2}$}

\date{}

\footnotetext[1]{
Department of Statistics and Operations Research, University of Santiago de Compostela (Spain).}
\footnotetext[2]{
Forest Research Centre, Technical University of Lisbon (Portugal).}
\footnotetext[3]{Corresponding author. e-mail: \href{mailto:eduardo.garcia@usc.es}{eduardo.garcia@usc.es}.}

\maketitle


\begin{abstract}
A nonparametric test for assessing the independence between a directional random variable (circular or spherical, as particular cases) and a linear one is proposed in this paper. The statistic is based on the squared distance between nonparametric kernel density estimates and its calibration is done by a permutation approach. The size and power characteristics of various variants of the test are investigated and compared with those for classical correlation-based tests of independence in an extensive simulation study. Finally, the best-performing variant of the new test is applied in the analysis of the relation between the orientation and size of Portuguese wildfires.
\end{abstract}

\begin{flushleft}
\end{flushleft}
\section{Introduction}
\label{sec:intro}

Characterization of wildfire orientation patterns at landscape scale has important management implications \citep{Moreira:2001p1105,Lloret:2002p1772,Moreira:2011p1102}. It has been  shown that landscape fuel reduction treatments will only be successful if strategically placed in order to intersect fire spread in the heading direction \citep{Finney:2001p457,Schmidt:2008p1386}. \\

\begin{figure}[h]
\centering
\includegraphics[width=0.32\textwidth]{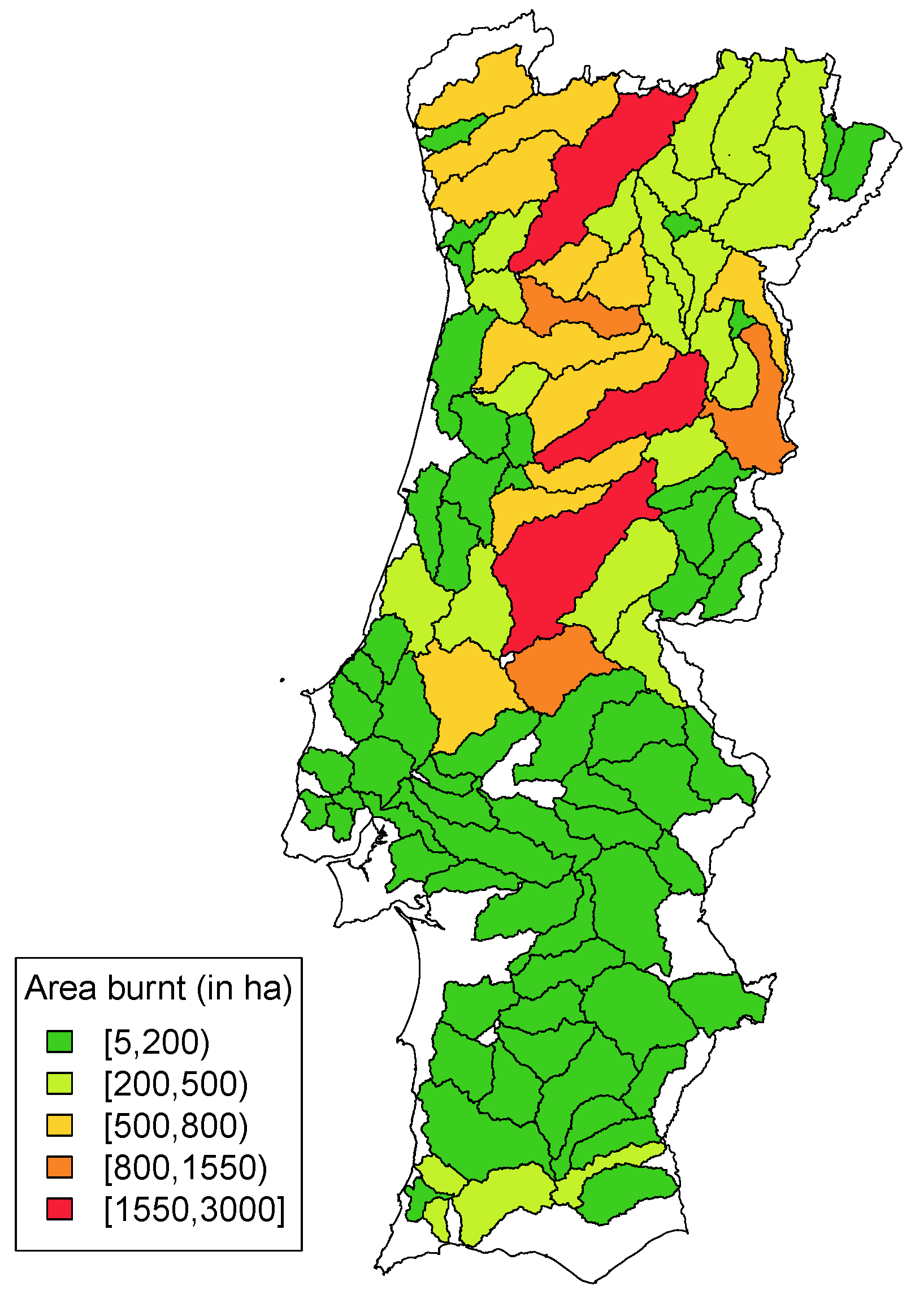}
\includegraphics[width=0.32\textwidth]{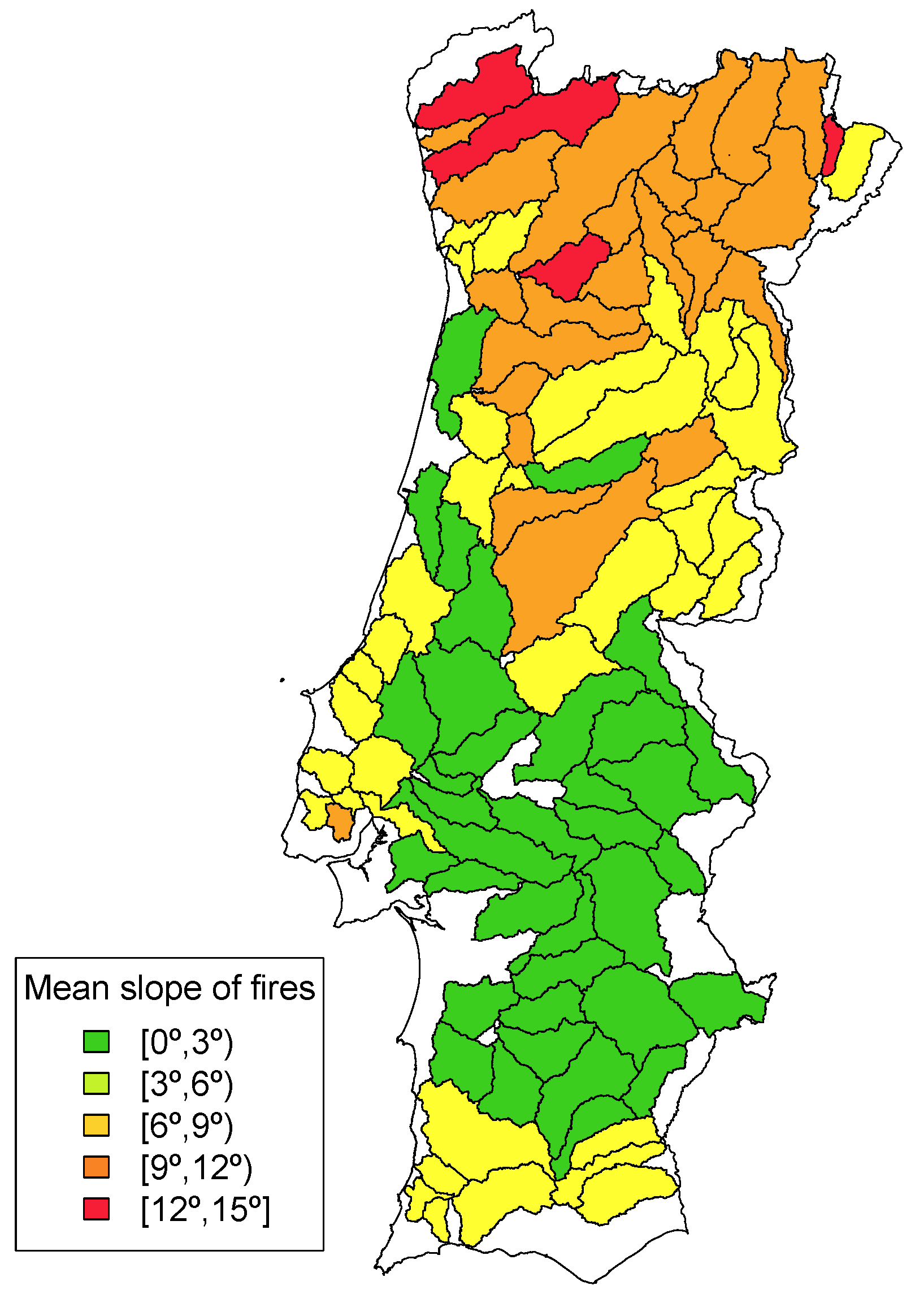}
\includegraphics[width=0.32\textwidth]{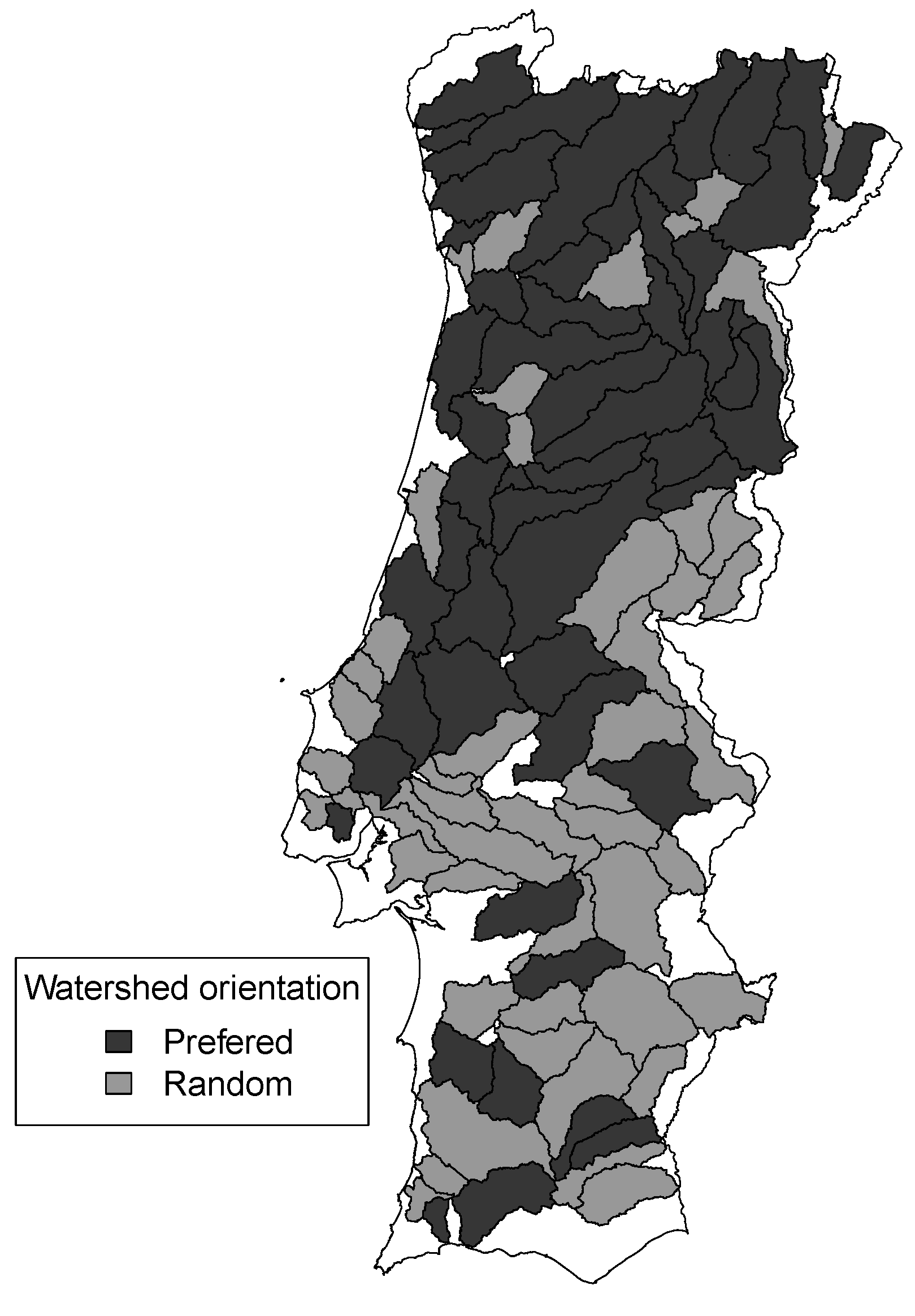}
\caption{\small Descriptive maps of wildfires in Portugal with the 102 watersheds delineated by \cite{Barros:2012p1135}. The left map shows the number of hectares burnt from fire perimeters associated with each watershed. Each fire perimeter is associated with the watershed that contains its centroid. The center map represents the mean slope of the fires of each watershed, where the slope is measured in degrees ($0^\circ$ stands for plain slope and $90^\circ$ for a vertical one). Finally, the right map shows watersheds where fires display preferential alignment according to \cite{Barros:2012p1135}.}
\label{fig3}       %
\end{figure}

\citet{Barros:2012p1135} assessed the existence of preferential fire perimeter orientation at watershed level, to support the spatial layout of fuelbreak networks. Their analysis identified clusters of watersheds where fire perimeters were preferentially aligned along the NE/\-SW and the SE/NW axes. Those watersheds included fire perimeters that together account for roughly $65\%$ of the overall burnt area in Portugal, over the period from 1975 to 2005, while in the remaining watersheds fire perimeters were randomly aligned. In Figure \ref{fig3}, some descriptive maps of the data of interest are displayed. The left plot shows the total area burnt in each watershed, whereas the middle plot represents the mean slope of the fires in each region. Finally, the right plot indicates which watersheds exhibit a preferred fire orientation, versus a random orientation, according to \cite{Barros:2012p1135}. The authors argued that spatial patterns of fire perimeter orientation found in the $31$-year dataset could be explained by dominant weather during the Portuguese fire season \citep{Pereira:2005p354}. However, given that fire perimeter orientation analysis is event-based (\textit{i.e.}, it is based on the orientation of each fire event) all perimeters are treated equally independently of their size. In this paper, a test for assessing independence between wildfire size and orientation is presented, complementing the work of \citet{Barros:2012p1135}. Furthermore, orientation of the wildfire will be considered in two-dimensional and three-dimensional spaces. \\

Spatial characterization of a wildfire, by means of its main orientation, and the associated burnt area, must be handled by non-standard statistical approaches, given the special nature of fire orientation. Specifically, it can be measured as an angle in the plane (two-dimensional orientation) or as a pair of angles identifying a direction in the three-dimensional sphere, if the main slope of the wildfire is taken into account. Hence, appropriate methods for handling circular and, more generally, directional data must be considered, jointly with suitable combinations of directional and linear techniques.\\

The analysis of the relation between directional and linear variables has been classically approached through the construction of circular-linear correlation co\-ef\-fi\-cients. The adaptation of the classical linear correlation coefficient to the circular-linear setting was introduced by \cite{Mardia1976} and \cite{Johnson1977} and further studied by \cite{Liddell1978}, who obtained its exact distribution under certain parametric assumptions. For the circular-linear case, a rank-based test of association was also proposed by \cite{Mardia1976}, who derived its asymptotic distribution. Later, \cite{Fisher1981} adapted Kendall's $\tau$ as a measure of circular-linear association based on the notion of concordance in the cylinder. To the best of the authors' knowledge, these three tests are the only available for testing the independence in directional-linear variables. As they are based on correlation coefficients, these tests are only powerful against deviations in the conditional expectation that can be measured by the corresponding coefficient. As a consequence, none of these tests are able to capture all possible types of dependence, neither for the conditional expectation nor for more complex types of dependence.

From a different perspective, circular and linear variables can also be jointly modeled by the construction of circular-linear distributions. \cite{Johnson1978} introduced a method for deriving circular-linear densities with specified marginals. A new family of cir\-cu\-lar\--linear distributions based on nonnegative trigonometric sums, which proved to be more flexible in capturing the data structure, was proposed by \cite{Fern'andez-Dur'an2007}, adapting the method by \cite{Johnson1978}. More recently, \cite{Garcia-Portugues:so2} exploited the copula representation of the \cite{Johnson1978} family, allowing for a completely nonparametric estimator, which was applied to analyze SO$_2$ concentrations and wind direction. Nevertheless, the aforementioned methods are designed for the circular-linear case, whereas in our context, a more general tool for handling directional-linear relations is needed, provided that wildfire orientation may be reported in two or three dimensions.\\

In this paper, the assessment of the relation between a directional (circular or spherical, as particular cases) and a linear variable is approached through the construction of a formal test to check directional-linear independence. Inspired by the ideas of \cite{Rosenblatt1975} and \cite{Rosenblatt1992} for the linear setting (see also \cite{Ahmad1997}), the proposed test statistic is based on a nonparametric directional-linear kernel density estimator and an $\mathcal{L}_2$ distance is taken as a discrepancy measure between the joint estimator and the one constructed under the independence hypothesis. The new test presents some interesting advantages: it is designed in a general fashion for directional variables of all dimensions and it is able to capture all kinds of deviations from independence by virtue the nonparametric density estimation. Besides, one gets a kernel density estimate as a spin-off, which provides further information about the form of dependence when independence is rejected.\\

The remainder of the paper is organized as follows. In Section \ref{sec:back}, some background to kernel density estimation, for linear, directional and directional-linear data is presented. Section \ref{sec:test} is devoted to the introduction of the test statistic, introducing a simplified version of the test and describing in detail its practical application. The finite sample performance of the test, in terms of size and power, is assessed through a simulation study for circular-linear and spherical-linear variables. Application to real data is provided in Section \ref{sec:realdata}, including data description and results, focusing on the assessment of independence between wildfire orientation and burnt area size in Portugal. Some discussion and final comments are given in Section \ref{sec:discussion}.

%
\section{Background to kernel density estimation}
\label{sec:back}

In the linear setting, the basic building block for the independence test introduced by \cite{Rosenblatt1975} is a kernel density estimator. Independence between two linear random variables is assessed through an $\mathcal{L}_2$ distance between a bidimensional kernel density estimator and the product of the marginal kernel density estimators. In order to extend such a procedure to the directional-linear case, kernel density estimation for linear, directional and directional-linear variables is required. A brief background on kernel density estimators will be provided in this\nolinebreak[4] section. 
\subsection{Linear kernel density estimation}
The well-known kernel density estimator for linear data was introduced by \cite{Rosenblatt1956} and \cite{Parzen1962}. Given a random sample $Z_1,\ldots,Z_n$ from a linear random variable $Z$ (\textit{i.e.} with support $\mathrm{supp}(Z)\subseteq\mathbb R$) with density $f$, the kernel density estimator at a point $z\in \R$ is defined as
\[
\hat f_g(z)=\frac{1}{ng}\sum_{i=1}^n K\lrp{\frac{z-Z_i}{g}},%
\]
where $K$ is a kernel function, usually a symmetric density about the origin, and $g>0$ is the smoothing or bandwidth parameter, which controls the roughness of the estimator. Properties of this estimator have been deeply studied (see \cite{Silverman1986} or \cite{Wand1995} for comprehensive reviews). It is also well known that the choice of kernel (normal, Epanechnikov, etc.) has little effect on the overall shape of the kernel density estimate. However, the bandwidth is a key tuning parameter: large values produce oversmoothed estimates of $f$, whereas small values provide undersmoothed curves. Comprehensive reviews on bandwidth selection are given in \cite{Cao1994}, \cite{Chiu1996} and \cite{Jones1996}, among others.

\subsection{Directional kernel density estimation}
Denote by $\bX$ a directional random variable with density $f$. The support of such a variable is the $q$-dimensional sphere, namely $\Om{q}=\big\{\bx\in\R^{q+1}:x^2_1+\cdots+x^2_{q+1}=1\big\}$, endowed with the Lebesgue measure in $\Om{q}$, that will be denoted by $\om{q}$. Therefore, a directional density is a nonnegative function that satisfies $\Iq{f(\bx)}{\bx}=1$.\\

The directional kernel density estimator was introduced by \cite{Hall1987} and \cite{Bai1988}. Given a random sample $\bX_1,\ldots,\bX_n$, of a directional variable $\bX$ with $\mathrm{supp}(\bX)\subseteq\Omega_q$ and density $f$, at a point $\bx\in \Omega_q$ the estimator is given by
\begin{align}
\hat f_h(\bx)=\frac{c_{h,q}(L)}{n}\sum_{i=1}^n L\lrp{\frac{1-\bx^T\bX_i}{h^2}},\label{kernel_directional}%
\end{align}
where $L$ is the directional kernel, $h>0$ is the bandwidth parameter and $c_{h,q}(L)$ is a normalizing constant depending on the kernel $L$, the bandwidth $h$ and the sphere dimension $q$. The scalar product of two vectors, $\bx$ and $\by$, is denoted by $\bx^T\by$, where $^T$ denotes the transpose operator. \\

A common choice for the directional kernel is $L(r)=e^{-r}$, $r\geq0$, also known as the von Mises kernel due to its relation with the von Mises-Fisher distribution (see \cite{Watson1983}). In a $q$-dimensional sphere, the von Mises density $\textrm{vM}(\bmu,\kappa)$ is given by
\begin{align}
f_{\textrm{vM}}(\bx;\bmu,\kappa)=C_q(\kappa) \exp{\lrb{\kappa\bx^T\bmu}},\quad C_q(\kappa)=\kappa^{\frac{q-1}{2}}\lrc{(2\pi)^{\frac{q+1}{2}}\mathcal{I}_{\frac{q-1}{2}}(\kappa)}^{-1},\label{dir:cq}
\end{align}
where $\bmu\in\Omega_q$ is the mean direction, $\kappa\geq0$ is the concentration parameter around the mean and $\mathcal{I}_\nu$ is the modified Bessel function of order $\nu$,
\begin{align*}
\mathcal{I}_\nu(z)=\frac{\lrp{\frac{z}{2}}^\nu}{\pi^{1/2}\Gamma\lrp{\nu+\frac{1}{2}}}\int_{-1}^1 (1-t^2)^{\nu-\frac{1}{2}}e^{zt}\,dt.
\end{align*}
For the von Mises kernel, the value of $c_{h,q}(L)$ is $C_q\lrp{1/h^2}\allowbreak e^{1/h^2}$ and the directional estimator (\ref{kernel_directional}) can be interpreted as a mixture of von Mises-Fisher densities:
\begin{align*}
\hat f_h(\bx)=\frac{1}{n}\sum_{i=1}^n f_{\textrm{vM}}\lrp{\bx;\bX_i,1/h^2}.
\end{align*}
Note that large values of $h$ provide a small concentration parameter, which results in a uniform model in the sphere, whereas small values of $h$ give high concentrations around the sample observations, providing an undersmoothed curve. Cross-validation rules based on Likelihood Cross Validation (LCV) and Least Squares Cross Validation (LSCV) for bandwidth selection were discussed by \cite{Hall1987}.

\subsection{Directional-linear kernel density estimation.}
Consider a directional-linear random variable, $(\bX,Z)$ with support $\mathrm{supp}(\bX,Z)\subseteq \Omega_q\times\mathbb{R}$ and joint density $f$. For the simple case of circular data ($q=1$), the support of the variable is the cylinder and, in general, the support is a multidimensional cylinder. Following the ideas in the previous sections for the linear and directional cases, given a random sample $\lrp{\bX_1,Z_1},\ldots,\lrp{\bX_n,Z_n}$, the directional-linear kernel density estimator at a point $(\bx,z)\in\Omega_q\times\mathbb R$ can be defined as
\begin{align}
\hat f_{h,g}(\bx,z)=\frac{c_{h,q}(L)}{ng}\sum_{i=1}^nLK\lp\frac{1-\bx^T\bX_i}{h^2},\frac{z-Z_i}{g}\rp,%
\label{dirlin}
\end{align}
where $LK$ is a directional-linear kernel, $g>0$ is the linear bandwidth parameter, $h>0$ is the directional bandwidth and $c_{h,q}(L)$ is the directional normalizing constant. The estimator (\ref{dirlin}) was introduced by \cite{Garcia-Portugues:dirlin}, who also studied its asymptotic properties\nopagebreak[4] in\nopagebreak[4] terms of bias and variance, and established its asymptotic normality. \\

A product kernel $LK(\cdot,\cdot)=L(\cdot)\times K(\cdot)$, specifically, the von Mises-normal kernel
\begin{align*}
LK(r,t)=e^{-r}\times\phi_1(t),\quad r\in[0,\infty),\,t\in\R,
\end{align*}
will be considered throughout this paper in order to simplify computations, where $\phi_\sigma$ denotes the density of a normal with zero mean and standard deviation $\sigma$. Similarly to the linear and directional kernel density estimators, a smoothing parameter (bidimensional, in this case) is involved in the estimator construction. The cross-validation procedures introduced by \cite{Hall1987} can be adapted to the directional-linear setting, yielding the following bandwidth selectors:
\begin{align*}
(h,g)_\mathrm{LCV}=&\,\arg\max_{h,g>0} \sum_{i=1}^n \log\hat f_{h,g}^{-i}(\bX_i,Z_i),\\
(h,g)_\mathrm{LSCV}=&\,\arg\max_{h,g>0} \lrc{2n^{-1}\sum_{i=1}^n\hat f_{h,g}^{-i}(\bX_i,Z_i) -\Iqr{\hat f_{h,g}(\bx,z)^2}{\bx}{z}},
\end{align*}
where $f_{h,g}^{-i}$ represents the kernel density estimator computed without the $i$-th datum.

\section{A test for directional-linear independence}
\label{sec:test}

The new test statistic for assessing independence between a directional and a linear variable is described in this section. 

\subsection{The test statistic}
\label{subsec:teststat}

Consider the joint directional-linear density $f_{(\bX,Z)}$ for the variable $(\bX,Z)$. $f_\bX$ and $f_Z$ denote the directional and linear marginal densities, respectively. The null hypothesis of independence between both components can be stated as
\begin{align*}
H_0: f_{(\bX,Z)}(\bx,z)=f_\bX(\bx)f_Z(z),\quad \forall (\bx,z)\in\Om{q}\times\R
\end{align*}
and the alternative hypothesis as
\begin{align*}
H_a: f_{(\bX,Z)}(\bx,z)\neq f_\bX(\bx)f_Z(z),\quad \text{for any }(\bx,z)\in\Om{q}\times\R.
\end{align*}

Following the idea of \cite{Rosenblatt1975}, a natural statistic to test $H_0$ arises from considering the $\mathcal{L}_2$ distance between the nonparametric estimation of the joint density $f_{(\bX,Z)}$ by the directional-linear kernel estimator (\ref{dirlin}), denoted by $\hat f_{(\bX,Z);h,g}$, and the nonparametric estimation of $f_{(\bX,Z)}$ under $H_0$, given by the product of the marginal directional and linear kernel estimators, denoted by $\hat f_{\bX;h}$ and $\hat f_{Z;g}$, respectively. We therefore propose the following test statistic:
\begin{align}
T_n&=\Delta_2\Big(\hat f_{(\bX,Z);h,g},\hat f_{\bX;h}\hat f_{Z;g}\Big),\label{test}
\end{align}
where $\Delta_2$ stands for the squared $\mathcal{L}_2$ distance in $\Om{q}\times\R$ between two functions $f_1$ and $f_2$:
\begin{align}
\Delta_2(f_1,f_2)&=\int_{\Om{q}\times\R}\lrp{f_1(\bx,z)-f_2(\bx,z)}^2\,\om{q}(d\bx)\,dz.\nonumber
\end{align}
The test statistic depends on a pair of bandwidths $(h,g)$, which is used for the directional-linear estimator, and whose components are also considered for the marginal directional and linear kernel density estimators. Under the null hypothesis of independence, $H_0$, it holds that $\mathbb{E}\big[\hat f_{(\bX,Z);h,g}(\bx,z)\big]=\mathbb{E}\big[\hat f_{\bX;h}(\bx)\big]\allowbreak\mathbb{E}\big[\hat f_{Z;g}(z)\big]$.\\

Asymptotic properties of (\ref{test}) have been studied by \cite{Garcia-Portugues:clt}, who proved its asymptotic normality under independence, but with a slow rate of convergence that does not encourage its use in practice. For that reason, a calibration mechanism will be needed for the practical application of the test.\\

In addition, the construction of $T_n$ requires the calculation of an integral over $\Omega_q\times\mathbb R$, which may pose computational problems since it involves the calculation of several nested integrals. However, if the kernel estimators are obtained using von Mises and normal kernels, then an easy to compute expression for $T_n$ can be obtained, as stated in the following lemma.

\begin{lem}
\label{lem:1}
If the kernel estimators involved in (\ref{test}), obtained from a random sample\linebreak $\{(\bX_i,Z_i)\}_{i=1}^n$ of $(\bX,Z)$, are constructed with von Mises and normal kernels, the following expression for $T_n$ holds:
\begin{align}
T_n=\mathbf{1}_n\bigg(\frac{1}{n^2}\mathbf{\Psi}(h)\circ\mathbf{\Omega}(g)-\frac{2}{n^3}\mathbf{\Psi}(h)\mathbf{\Omega}(g)+\frac{1}{n^4}\mathbf{\Psi}(h)\mathbf{1}_n\mathbf{1}_n^T\mathbf{\Omega}(g)\bigg)\mathbf{1}_n^T,\label{test_simple}
\end{align}
where $\circ$ denotes the Hadamard product and $\mathbf{\Psi}(h)$ and $\mathbf{\Omega}(g)$ are $n\times n$ matrices given by
\begin{align*}
\mathbf{\Psi}(h)=\Bigg(\frac{C_q\lrp{1/h^2}^2}{C_q\lrp{\norm{\bX_i+\bX_j}/h^2}}\Bigg)_{ij},\quad\mathbf{\Omega}(g)=\lrp{\phi_{\sqrt{2}g}\lrp{Z_i-Z_j}}_{ij},
\end{align*}
where $\mathbf{1}_n$ is a vector of $n$ ones and $C_q$ is the normalizing function (\ref{dir:cq}). %
\end{lem}

The proof of this result can be seen in Appendix \ref{sec:appendixA}. Note that expression (\ref{test_simple}) for $T_n$ only requires matrix operations. This will be the expression used for computing the test statistic. It should also be noted that the effect of the dimension $q$ appears only in the definition of $C_q$ and in $\norm{\bX_i+\bX_j}$, and both are easily scalable for large $q$. Thus, an important advantage of (\ref{test_simple}) is that computing requirements are similar for different dimensions $q$, something which is not the case if (\ref{test}) is employed with numerical integration.

\subsection{Calibration of the test}
\label{subsec:testpractise}

The null hypothesis of independence is stated in a nonparametric way, which determines the resampling methods used for calibration. However, as the null hypothesis is of a non-interaction kind, a permutation approach (which is not at all foreign to hypothesis testing) seems a reliable option. If $\lrb{\lrp{\bX_i,Z_i}}_{i=1}^n$ is a random sample from the di\-rec\-tion\-al-linear variable $\lrp{\bX,Z}$ and $\sigma$  is a random permutation of $n$ elements, then $\big\{\big(\bX_i,Z_{\sigma(i)}\big)\big\}_{i=1}^n$, represents the resulting $\sigma$-permuted sample. $T_n^\sigma$ denotes the test statistic computed from the $\sigma$-permuted random sample. Under the assumption of independence between the directional and linear components, it is reasonable to expect that the distribution of $T_n$ is similar to the distribution of $T_n^\sigma$, which can be easily approximated by Monte Carlo methods.\\

In addition to its simplicity, the main advantage of the use of permutations is its easy implementation using Lemma \ref{lem:1}, as it is possible to reuse the computation of the matrices $\mathbf{\Psi}(h)$ and $\mathbf{\Omega}(g)$ needed for $T_n$ to compute a $\sigma$-permuted statistic $T_n^{\sigma}$. In virtue of expression (\ref{test_simple}) and the definition of $T_n^\sigma$, the $\sigma$-permuted test statistic is given by
\begin{align*}
T_n^\sigma=\mathbf{1}_n\bigg(\frac{1}{n^2}\mathbf{\Psi}(h)\circ\mathbf{\Omega^\sigma}(g)-\frac{2}{n^3}\mathbf{\Psi}(h)\mathbf{\Omega^\sigma}(g)+\frac{1}{n^4}\mathbf{\Psi}(h)\mathbf{1}_n\mathbf{1}_n^T\mathbf{\Omega^\sigma}(g)\bigg)\mathbf{1}_n^T,
\end{align*}
where the $ij$-th entry of the matrix $\mathbf{\Omega^\sigma}(g)$ is the $\sigma(i)\sigma(j)$-entry of $\mathbf{\Omega}(g)$. For the computation of $\mathbf{\Psi}(h)$ and $\mathbf{\Omega}(g)$, symmetry properties reduce the number of computations and can also be used to optimize the products $\mathbf{\Psi}(h)\circ\mathbf{\Omega^\sigma}(g)$ and $\mathbf{\Psi}(h)\mathbf{\Omega^\sigma}(g)$. The last addend of $T_n^\sigma$ is the same as that of $T_n$ and there is no need to recompute it. The testing procedure can be summarized in the following algorithm.

\begin{algo}
\label{algo:3}
Let $\lrb{(\bX_i,Z_i)}_{i=1}^n$ be a random sample from a directional-linear variable $(\bX,Z)$.
\begin{enumerate}[label=\textit{\roman{*}}., ref=\textit{\roman{*}}]
\item Obtain a suitable pair of bandwidths $(h,g)$.\label{algo:3:1}
\item Compute the observed value of $T_n$ from (\ref{test_simple}), with kernel density estimators taking bandwidths $(h,g)$.\label{algo:3:2}%
\item Permutation calibration. For $b=1,\ldots, B\leq n!$, compute $T_n^{\sigma_b}$ with bandwidths $(h,g)$ for a random permutation $\sigma_b$.\label{algo:3:3}
\item Approximate the $p$-value by $\#\big\{T_n\leq T_n^{\sigma_b}\big\}\big/B$, where $\#$ denotes the cardinal of the set.\label{algo:3:4}
\end{enumerate}
\end{algo}

In steps \ref{algo:3:2} and \ref{algo:3:3}, a pair of bandwidths must be chosen. For the directional-linear case, as commented in Section \ref{sec:back}, cross-validation bandwidths, namely $(h,g)_\mathrm{LCV}$ and $(h,g)_\mathrm{LSCV}$, can be considered. However, as usually happens with cross-val\-i\-datory bandwidths, these selectors tend to provide undersmoothed estimators, something which a priori is not desirable as introduces a substantial variability in the statistic $T_n$.\\

To mitigate this problem, a more sophisticated band\-width selector will be introduced. Considering the von Mises-normal kernel, the bootstrap version for the Mean Integrated Squared Error (MISE) of the directional-linear kernel density estimator (\ref{dirlin}) was derived by \cite{Garcia-Portugues:dirlin}:
\begin{align*}
\mathrm{MISE}_{h_p,g_p}^\ast\lrp{h,g}=&\,\Big(C_q(1/h^2)^2C_q(2/h^2)^{-1}2\pi^\frac{1}{2}gn\Big)^{-1}\\
&+n^{-2}\mathbf{1}_n\big[(1-n^{-1})\mathbf{\Psi_2^\ast}(h)\circ\mathbf{\Omega_2^\ast}(g)-2\mathbf{\Psi_1^\ast}(h)\circ\mathbf{\Omega_1^\ast}(g)+\mathbf{\mathbf{\Psi}_0^\ast}\circ\mathbf{\Omega_0^\ast}\big]\mathbf{1}_n^T,
\end{align*}
where matrices $\mathbf{\Psi^\ast_a}(h)$ and $\mathbf{\Omega^\ast_a}(g)$, $a=0,1,2$ are
\begin{align*}
\mathbf{\Psi_0^*}=&\,\lrp{\frac{C_q(1/h_p^2)^2}{C_q\big(||\bX_i+\bX_j||/h_p^2\big)}}_{ij},\quad\mathbf{\Psi_1^*}(h)= \lrp{\Iq{\frac{C_q(1/h^2)C_q(1/h_p^2)^2e^{\bx^T\bX_j/h_p^2}}{C_q\big(||\bx/h^2+\bX_i/h_p^2||\big)}}{\bx}\!}_{\!ij},\\
\mathbf{\Psi_2^*}(h)=&\, \Bigg(\Iq{\frac{C_q(1/h^2)^2C_q(1/h_p^2)^2}{C_q\big(||\bx/h^2+\bX_i/h_p^2||\big)C_q\big(||\bx/h^2+\bX_j/h_p^2||\big)}}{\bx}\Bigg)_{ij},\\
\mathbf{\Omega^*_0}=&\,\lrp{\phi_{\sqrt{2}g_p}(Z_i-Z_j)}_{ij},\quad\mathbf{\Omega^*_a}(g)=\lrp{\phi_{\sigma_{a,g}}(Z_i-Z_j)}_{ij},
\end{align*}
with $\sigma_{a,g}=\big(ag^2+2g_p^2\big)^\frac{1}{2}$, $a=1,2$, and $(h_p,g_p)$ a given pair of pilot bandwidths. Then, the estimation bandwidths are obtained as
\begin{align*}
(h,g)_{bo}=\arg\min_{h,g>0} \mathrm{MISE}_{h_p,g_p}^\ast\lrp{h,g}.
\end{align*}
The choice of $(h_p,g_p)$ is needed in order to compute $(h,g)_{bo}$. This must be done by a joint criterion for two important reasons. Firstly, to avoid the predominance of smoothing in one component that may dominate the other (this could happen, for example, if the directional variable is uniform, as in that case the optimal bandwidth tends to infinity). Secondly, to obtain a test with more power against deviations from independence. Based on these comments, a new bandwidth selector, named Bootstrap Likelihood Cross Validation (BLCV), is introduced:
\begin{align*}
(h,g)_{\mathrm{BLCV}}=\arg\min_{h,g>0} \mathrm{MISE}_{(h,g)_\mathrm{MLCV}}^\ast\lrp{h,g},
\end{align*}
where the pair of bandwidths $(h,g)_\mathrm{MLCV}$ are obtained by enlarging the order of $(h,g)_\mathrm{LCV}$ to be of the kind $\big(\mathcal O\big(n^{-1/(6+q)}\big), \mathcal O\big(n^{-1/7}\big)\big)$, the order that one would expect for a pair of directional-linear pilot bandwidths. For the linear component, this can be seen in the paper by \cite{Cao1993}, where the pilot bandwidth is proved to be $g_p=\mathcal O\big(n^{-1/7}\big)$, larger than the order of the optimal estimation bandwidth, $n^{-1/5}$. For the directional case there is no pilot bandwidth available, but considering that the order of the optimal estimation bandwidth is $n^{-1/(4+q)}$ \citep{Garcia-Portugues:dirlin}, then a plausible conjecture is $h_p=\mathcal O\big(n^{-1/(6+q)}\big)$.

\subsection{Simulation study}
\label{subsec:simus}

Six different directional-linear models were considered in the simulation study. The models are indexed by a $\delta$ parameter that measures the degree of deviation from the independence, where $\delta=0$ represents independence and $\delta>0$ accounts for different degrees of dependence. The models show three kind of possible deviations from the independence: first order deviations, that is, deviations in the conditional expectation (M1, M2 and M3); second order deviations or conditional variance deviations (M4 and M5) and first and second order deviations (M6). In order to clarify notation, $\phi(\cdot;m,\sigma)$ and $f_{\mathcal{LN}}(\cdot;m,\sigma)$ represent the density of a normal and a log-normal, with mean/log-scale $m$ and standard deviation/shape $\sigma$. Notation $\mathbf{0}_q$ represents a vector of $q$ zeros.
\begin{itemize}
\item[M1.] $f_1(\bx,z)=\phi\big(z;\delta(2+\bx^T\bmu),\sigma\big)\times f_{\textrm{vM}}(\bx;\bmu,\kappa)$, with $\bmu=(\mathbf{0}_q,1)$, $\kappa=1$ and $\sigma=1$.

\item[M2.] $f_2(\bx,z)=f_{\mathcal{LN}}\big(z;\delta(1+(\bx^T\bmu)^2),\sigma\big)\times f_{\textrm{vM}}(\bx;\bmu,\kappa)$, with $\bmu=(-1,\mathbf{0}_q)$, $\kappa=0$ and $\sigma=\frac{1}{4}$.

\item[M3.] $f_3(\bx,z)=\big[rf_{\mathcal{LN}}\big(z;\delta(1+(\bx^T\bmu_1)^3),\sigma_1\big)+(1-r)\phi(z;m,\sigma_2)\big] \times \big[pf_{\textrm{vM}}(\bx;\bmu_1,\kappa_1)+(1-p)$ $\times f_{\textrm{vM}}(\bx;\bmu_2,\kappa_2)\big]$, with $\bmu_1=(\mathbf{0}_q,1)$, $\bmu_2=(\mathbf{0}_q,-1)$, $\kappa_1=2$, $\kappa_2=1$, $p=\frac{3}{4}$, $\sigma_1=\sigma_2=\frac{1}{4}$, $m=1$ and $r=\frac{1}{4}$.

\item[M4.] $f_4(\bx,z)=\phi\big(z;m,\frac{1}{4}+\delta(1-(\bx^T\bmu_r)^3)\big)\times f_{\textrm{vM}}(\bx;\bmu,\kappa)$, with $\bmu=(\mathbf{0}_q,1)$, $\kappa=1$, $\bmu_r=(-1,\mathbf{0}_q)$ and $m=0$.

\item[M5.] $f_5(\bx,z)=f_{\mathcal{LN}}\big(z;m,(5+\delta\bx^T(3\bmu_2-\bmu_1))^{-1}\big)\times \lrc{pf_{\textrm{vM}}(\bx;\bmu_1,\kappa_1)+(1-p)f_{\textrm{vM}}(\bx;\bmu_2,\kappa_2)}$, with $\bmu_1\allowbreak=(\mathbf{0}_q,1)$, $\bmu_2=(\mathbf{0}_q,-1)$, $\kappa_1=\kappa_2=2$, $p=\frac{1}{2}$ and $m=0$.

\item[M6.] $f_6(\bx,z)=\big[rf_{\mathcal{LN}}(z;m,\sigma)+(1-r)\phi\big(z;\delta(2+\bx^T\bmu_r),\frac{1}{4}+\delta(\bx^T\bmu_r)^2\big) \big] \times f_{\textrm{vM}}(\bx;\bmu,\kappa)$, with $\bmu=(\mathbf{0}_q,1)$, $\bmu_r=(-1,\mathbf{0}_q)$, $\kappa=1$, $p=\frac{3}{4}$, $\sigma=\frac{1}{2}$, $m=0$ and $r=\frac{3}{4}$.
\end{itemize}
The choice of the models was done in order to capture situations with heteroskedasticity, skewness in the linear component and different types of von Mises mixtures in the directional component.\\

\begin{figure}[ht]
\centering
\includegraphics[scale=0.45]{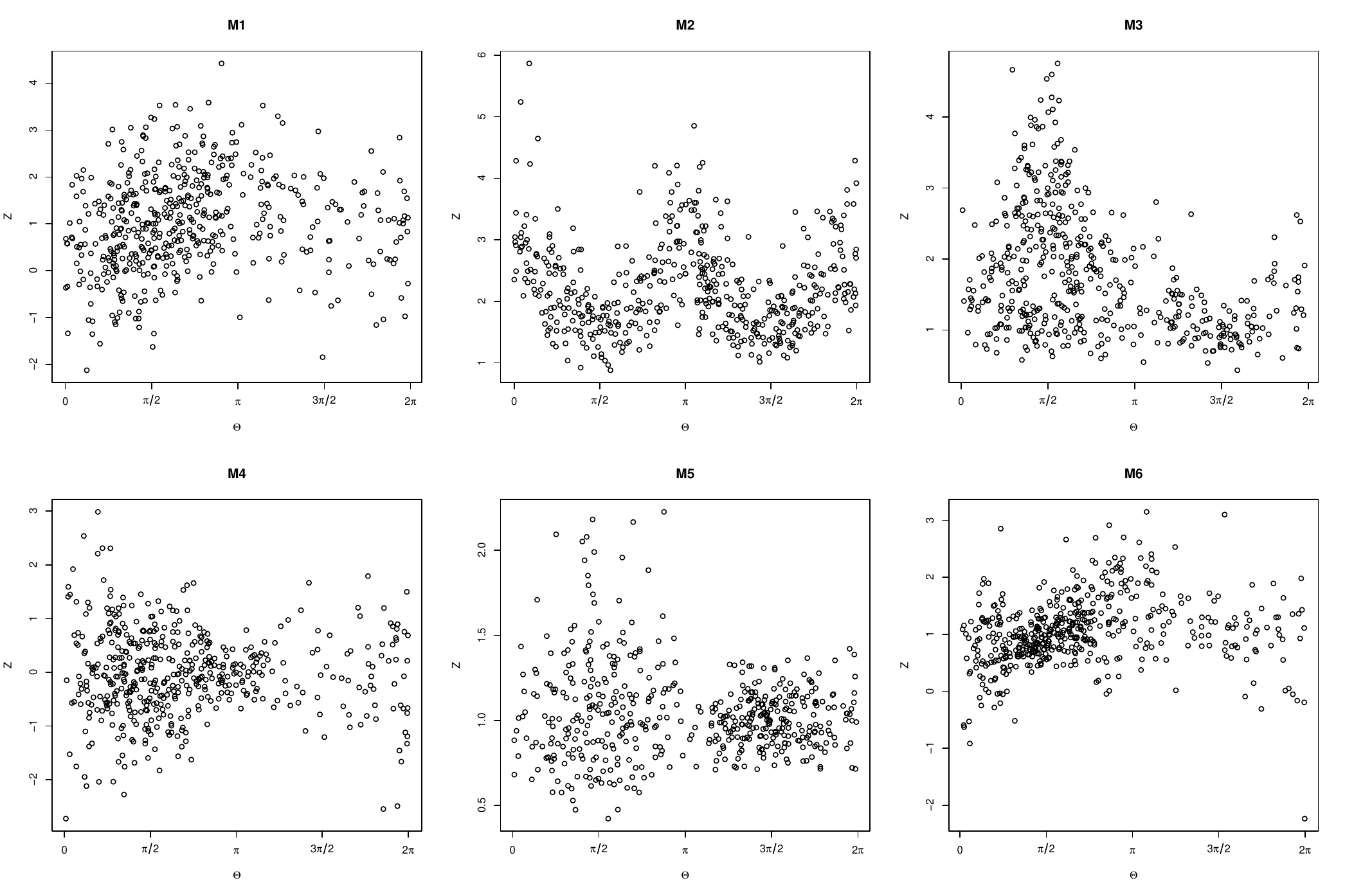}
\caption{\small Random samples of $n=500$ points for the simulation models in the circular-linear case, with $\delta=0.50$ (situation with dependence). From left to right and up to down, M1 to M6. M1, M2 and M3 present a deviation from the independence in terms of the conditional expectation; M4 and M5 account for a deviation in terms of the conditional variance and M6 includes deviations both in conditional expectation and variance. }
\label{fig2}       %
\end{figure}

For the proposed models, different deviations from independence have been considered, by setting $\delta=0,\allowbreak 0.25,0.50$. The proposed test statistic has been computed for all the models and sample sizes $n=50$, $100$, $200$, $500$, $1000$. The new test based on the permutation resampling described in Algorithm \ref{algo:3}, depending on the bandwidth choice, is denoted by $T_n^\mathrm{LCV}$ and $T_n^\mathrm{BLCV}$. The number of permutations considered was $B=1000$ and the number of Monte Carlo replicates was $M=1000$. Both the circular-linear and spherical-linear cases were explored. For the circular-linear case, the test was compared with the three tests available for circular-linear association, described as follows:
\begin{itemize}
\item Circular-linear correlation coefficient from \cite{Mardia1976} and \cite{Johnson1977}, denoted by $R_n^2$.
\item Rank circular-linear correlation coefficient from \cite{Mardia1976}, denoted by $U_n$.
\item $\lambda_{4n}$ measure of cylindrical association of \cite{Fisher1981}, implemented with its incomplete version $\lambda_{4n}^*$ considering $m=5000$ random $4$-tuples.
\end{itemize}

Although there exists an exact distribution for $R_n^2$ under certain normality assumptions on the linear response and asymptotic distributions for $U_n$ and $\lambda_{4n}^*$, for a fair comparison, the calibration of these tests has also been done by permutations ($B=1000$). The exact and asymptotic distributions for $R_n^2$ and $U_n$ were also tried instead of the permutation approach, providing empirical levels and powers quite similar to the ones based on permutations.\\

The proportion of rejections under $H_{k,\delta}$ (for model number $k$ with $\delta$ deviation) is reported in Tables \ref{tab1} and \ref{tab1b}, for the circular-linear and spherical-linear cases, with different sample sizes. In the circular-linear case, the empirical size is close to the nominal level for all the competing tests. The $T_n^\mathrm{LCV}$ test for this case shows in general a satisfactory behavior under the null hypothesis, except for some cases in M1, M4 and M5, where the test tends to reject the null hypothesis more than expected. This is mostly corrected by $T_n^\mathrm{BLCV}$, with a decrease of power with respect to $T_n^\mathrm{LCV}$ in M2. For the spherical-linear case, the improvement in size approximation $T_n^\mathrm{BLCV}$ is notable, specially for small sample sizes. If the tests maintain the nominal significance level of $5\%$, it is expected that approximately $95\%$ of the observed proportions of rejections under the null hypothesis\nopagebreak[4] (\textit{i.e.} when $\delta=0$) to lie within the interval $(0.036,0.064)$ to three decimal places.\\

\begin{table}[htpb!]
\centering
\scriptsize
\vspace*{-0.35cm}
\begin{tabular}{ll| >{\centering\arraybackslash}m{1.0cm}>{\centering\arraybackslash}m{1.0cm}>{\centering\arraybackslash}m{1.0cm}|>{\centering\arraybackslash}m{1.0cm}>{\centering\arraybackslash}m{1.0cm}||>{\centering\arraybackslash}m{1.0cm}>{\centering\arraybackslash}m{1.0cm}}
\toprule\toprule
\multirow{2}{*}{$n$} & \multirow{2}{*}{Model} & \multicolumn{5}{c||}{Circular-linear} & \multicolumn{2}{c}{Spherical-linear} \\\cmidrule(lr){3-7} \cmidrule(lr){7-9}
 &  & $R^2_n$ &$U_n$&$\lambda^*_{4n}$&$T_n^{\mathrm{LCV}}$&$T_n^{\mathrm{BLCV}}$&$T_n^{\mathrm{LCV}}$&$T_n^{\mathrm{BLCV}}$\\
\midrule
$50$
& $H_{1,0.00}$    &$0.047$ & $0.040$ & $0.047$ & {$0.050$} & $0.047$ & $0.059$ & {$0.045$} \\
& $H_{2,0.00}$    &{$0.051$} & $0.044$ & $0.042$ & $0.055$ & $0.052$ & $0.057$ & {$0.052$} \\
& $H_{3,0.00}$    &$0.047$ & $0.045$ & {$0.051$} & $0.059$ & $0.054$ & {$0.051$} & $0.048$ \\
& $H_{4,0.00}$    &$0.047$ & $0.040$ & $0.047$ & {$0.050$} & $0.046$ & $0.059$ & {$0.046$} \\
& $H_{5,0.00}$    &$0.042$ & $0.045$ & $0.053$ & $0.055$ & $0.047$  & $0.070$ & {$0.057$} \\
& $H_{6,0.00}$    &$0.058$ & $0.065$ & $0.062$ & $0.057$ & {$0.055$} & $0.065$ & {$0.054$} \\\midrule
& $H_{1,0.25}$    &{$0.162$} & $0.120$ & $0.092$ & $0.132$ & $0.139$ & {$0.099$} & $0.094$ \\
& $H_{2,0.25}$    &$0.055$ & $0.074$ & $0.088$ & {$0.143$} & $0.071$ & {$0.072$} & $0.050$ \\
& $H_{3,0.25}$    &$0.535$ & $0.538$ & $0.365$ & $0.511$ & {$0.543$} & $0.238$ & {$0.246$} \\
& $H_{4,0.25}$    &$0.051$ & $0.044$ & $0.067$ & {$0.239$} & $0.234$ & {$0.103$} & $0.097$ \\
& $H_{5,0.25}$    &$0.046$ & $0.049$ & $0.059$ & $0.128$ & $0.121$  & {$0.110$} & $0.094$ \\
& $H_{6,0.25}$    &$0.354$ & $0.332$ & $0.239$ & {$0.436$} & $0.432$ & {$0.284$} & $0.275$ \\\midrule
& $H_{1,0.50}$    &{$0.512$} & $0.412$ & $0.235$ & $0.378$ & $0.421$ & $0.231$ & {$0.253$} \\
& $H_{2,0.50}$    &$0.054$ & $0.124$ & $0.261$ & {$0.633$} & $0.291$ & {$0.219$} & $0.078$ \\
& $H_{3,0.50}$    &$0.925$ & $0.845$ & $0.734$ & $0.929$ & {$0.949$} & $0.662$ & {$0.666$} \\
& $H_{4,0.50}$    &$0.058$ & $0.050$ & $0.081$ & {$0.424$} & $0.420$ & {$0.149$} & $0.139$ \\
& $H_{5,0.50}$    &$0.055$ & $0.059$ & $0.094$ & {$0.501$} & $0.491$ & $0.320$ & $0.298$  \\
& $H_{6,0.50}$    &{$0.782$} & $0.706$ & $0.536$ & $0.754$ & $0.756$ & {$0.556$} & $0.540$ \\\midrule
$100$
& $H_{1,0.00}$    &{$0.052$} & $0.054$ & $0.063$ & $0.068$ & $0.061$ & $0.072$ & $0.068$  \\
& $H_{2,0.00}$    &$0.044$ & $0.046$ & {$0.052$} & $0.053$ & {$0.048$} & {$0.051$} & $0.055$  \\
& $H_{3,0.00}$    &$0.047$ & {$0.050$} & $0.046$ & $0.061$ & $0.054$ & $0.064$ & {$0.049$} \\
& $H_{4,0.00}$    &{$0.052$} & $0.054$ & $0.063$ & $0.067$ & $0.060$ & $0.072$ & $0.071$  \\
& $H_{5,0.00}$    &$0.056$ & {$0.050$} & $0.057$ & $0.073$ & $0.063$ & $0.074$ & {$0.063$} \\
& $H_{6,0.00}$    &$0.046$ & $0.046$ & {$0.050$} & $0.062$ & $0.059$ & $0.077$ & $0.071$ \\\midrule
& $H_{1,0.25}$    &{$0.291$} & $0.227$ & $0.102$ & $0.211$ & $0.213$ & $0.155$ & {$0.163$} \\
& $H_{2,0.25}$    &$0.051$ & $0.073$ & $0.092$ & {$0.263$} & $0.094$ & {$0.114$} & $0.067$ \\
& $H_{3,0.25}$    &{$0.889$} & $0.851$ & $0.407$ & $0.805$ & $0.849$ & $0.487$ & {$0.500$} \\
& $H_{4,0.25}$    &$0.060$ & $0.049$ & $0.074$ & $0.478$ & {$0.484$} & {$0.222$} & $0.219$ \\
& $H_{5,0.25}$    &$0.063$ & $0.050$ & $0.067$ & {$0.260$} & $0.251$ & $0.171$ & $0.171$  \\
& $H_{6,0.25}$    &$0.547$ & $0.574$ & $0.283$ & {$0.720$} & $0.718$ & {$0.492$} & $0.479$ \\\midrule
& $H_{1,0.50}$    &{$0.847$} & $0.721$ & $0.290$ & $0.669$ & $0.718$ & $0.416$ & {$0.460$} \\
& $H_{2,0.50}$    &$0.053$ & $0.122$ & $0.279$ & {$0.940$} & $0.660$ & {$0.530$} & $0.123$ \\
& $H_{3,0.50}$    &{$1.000$} & $0.997$ & $0.872$ & $0.999$ & $0.999$ & $0.942$ & $0.957$  \\
& $H_{4,0.50}$    &$0.058$ & $0.053$ & $0.103$ & $0.784$ & {$0.803$} & $0.341$ & {$0.355$} \\
& $H_{5,0.50}$    &$0.083$ & $0.056$ & $0.107$ & $0.836$ & {$0.860$} & $0.602$ & {$0.630$}  \\
& $H_{6,0.50}$    &$0.965$ & $0.951$ & $0.642$ & {$0.968$} & $0.967$ & {$0.864$} & $0.845$ \\\midrule
$200$
& $H_{1,0.00}$    &{$0.049$} & $0.056$ & $0.064$ & $0.057$ & $0.054$  & $0.065$ & $0.060$   \\
& $H_{2,0.00}$    &$0.055$ & $0.063$ & $0.055$ & $0.053$ & $0.054$   & {$0.046$} & $0.041$  \\
& $H_{3,0.00}$    &$0.051$ & $0.054$ & $0.058$ & $0.053$ & {$0.050$}  & {$0.045$} & $0.042$  \\
& $H_{4,0.00}$    &{$0.049$} & $0.056$ & $0.064$ & $0.057$ & $0.053$  & $0.065$ & $0.057$   \\
& $H_{5,0.00}$    &{$0.049$} & $0.056$ & $0.043$ & $0.066$ & $0.063$  & {$0.060$} & $0.061$  \\
& $H_{6,0.00}$    &$0.048$ & $0.059$ & $0.052$ & {$0.049$} & $0.046$  & $0.054$ & {$0.048$}  \\\midrule
& $H_{1,0.25}$    &{$0.529$} & $0.444$ & $0.099$ & $0.349$ & $0.373$  & $0.192$ & {$0.208$}  \\
& $H_{2,0.25}$    &$0.058$ & $0.081$ & $0.106$ & {$0.551$} & $0.154$  & {$0.178$} & $0.052$  \\
& $H_{3,0.25}$    &{$0.996$} & $0.995$ & $0.431$ & $0.980$ & $0.987$  & $0.795$ & {$0.818$}  \\
& $H_{4,0.25}$    &$0.054$ & $0.057$ & $0.085$ & $0.839$ & {$0.862$}  & $0.337$ & {$0.348$}  \\
& $H_{5,0.25}$    &$0.056$ & $0.052$ & $0.058$ & $0.459$ & {$0.487$}  & $0.303$ & $0.343$   \\
& $H_{6,0.25}$    &$0.830$ & $0.896$ & $0.277$ & {$0.974$} & $0.971$  & {$0.842$} & $0.830$  \\\midrule
& $H_{1,0.50}$    &{$0.982$} & $0.957$ & $0.299$ & $0.924$ & $0.940$  & $0.721$ & {$0.750$}  \\
& $H_{2,0.50}$    &$0.061$ & $0.145$ & $0.325$ & {$0.999$} & $0.967$  & {$0.899$} & $0.249$  \\
& $H_{3,0.50}$    &{$1.000$} & {$1.000$} & $0.913$ & {$1.000$} & {$1.000$}  & {$1.000$} & {$1.000$}  \\
& $H_{4,0.50}$    &$0.053$ & $0.058$ & $0.122$ & $0.981$ & {$0.984$}  & $0.595$ & {$0.618$}  \\
& $H_{5,0.50}$    &$0.124$ & $0.051$ & $0.105$ & $0.991$ & {$0.995$} & $0.921$ & {$0.950$}  \\
& $H_{6,0.50}$    &{$1.000$} & {$1.000$} & $0.691$ & {$1.000$} & {$1.000$} & {$0.994$} & $0.993$ \\\bottomrule\bottomrule
\end{tabular}
\caption{\small Proportion of rejections for the $R^2_n$, $U_n$, $\lambda^*_{4n}$, $T_n^{\mathrm{LCV}}$ and $T_n^{\mathrm{BLCV}}$ tests of independence for sample sizes $n=50,100,200$ for a nominal significance level of $5\%$. For the six different models the values of the deviation from independence parameter are $\delta=0$ (independence), $0.25$ and $0.50$. Each proportion was calculated using $B=1000$ permutations for each of $M=1000$ random samples of size $n$ simulated from the specified model. \label{tab1}}
\end{table}

\begin{table}[htpb!]
\centering
\scriptsize
\begin{tabular}{ll| >{\centering\arraybackslash}m{1.0cm}>{\centering\arraybackslash}m{1.0cm}>{\centering\arraybackslash}m{1.0cm}|>{\centering\arraybackslash}m{1.0cm}>{\centering\arraybackslash}m{1.0cm}||>{\centering\arraybackslash}m{1.0cm}>{\centering\arraybackslash}m{1.0cm}}
\toprule\toprule
\multirow{2}{*}{$n$} & \multirow{2}{*}{Model} & \multicolumn{5}{c||}{Circular-linear} & \multicolumn{2}{c}{Spherical-linear} \\\cmidrule(lr){3-7} \cmidrule(lr){7-9}
 &  & $R^2_n$ &$U_n$&$\lambda^*_{4n}$&$T_n^{\mathrm{LCV}}$&$T_n^{\mathrm{BLCV}}$&$T_n^{\mathrm{LCV}}$&$T_n^{\mathrm{BLCV}}$\\
\midrule
$500$
& $H_{1,0.00}$    &{$0.053$} & $0.060$ & $0.054$ & $0.069$ & $0.064$ &  {$0.055$} & {$0.045$} \\
& $H_{2,0.00}$    &$0.060$ & {$0.050$} & $0.053$ & $0.062$ & $0.055$ &  $0.046$ & {$0.048$} \\
& $H_{3,0.00}$    &$0.054$ & $0.059$ & $0.064$ & {$0.050$} & $0.044$ &  {$0.045$} & $0.042$ \\
& $H_{4,0.00}$    &{$0.053$} & $0.060$ & $0.054$ & $0.069$ & $0.062$ &  $0.056$ & {$0.050$} \\
& $H_{5,0.00}$    &$0.042$ & $0.038$ & $0.058$ & {$0.050$} & $0.047$ &  {$0.051$} & $0.059$ \\
& $H_{6,0.00}$    &$0.052$ & {$0.050$} & $0.053$ & $0.059$ & $0.062$ &  $0.059$ & {$0.055$} \\\midrule
& $H_{1,0.25}$    &{$0.916$} & $0.842$ & $0.088$ & $0.698$ & $0.727$ &  $0.422$ & {$0.447$} \\
& $H_{2,0.25}$    &$0.050$ & $0.073$ & $0.095$ & {$0.973$} & $0.511$ &  {$0.557$} & $0.073$ \\
& $H_{3,0.25}$    &{$1.000$} & {$1.000$} & $0.443$ & {$1.000$} & {$1.000$}  & $0.995$ & {$0.997$} \\
& $H_{4,0.25}$    &$0.057$ & $0.060$ & $0.077$ & {$0.999$} & {$0.999$} & $0.764$ & {$0.786$} \\
& $H_{5,0.25}$    &$0.080$ & $0.038$ & $0.068$ & $0.850$ & $0.865$ &  $0.679$ & {$0.750$} \\
& $H_{6,0.25}$    &$0.998$ & {$1.000$} & $0.263$ & {$1.000$} & {$1.000$}  & {$1.000$} & $0.998$ \\\midrule
& $H_{1,0.50}$    &{$1.000$} & {$1.000$} & $0.264$ & $0.998$ & $0.999$ & $0.982$ & {$0.986$} \\
& $H_{2,0.50}$    &$0.053$ & $0.125$ & $0.322$ & {$1.000$} & {$1.000$} & {$1.000$} & $0.910$ \\
& $H_{3,0.50}$    &{$1.000$} & {$1.000$} & $0.942$ & {$1.000$} & {$1.000$} & {$1.000$} & {$1.000$} \\
& $H_{4,0.50}$    &$0.064$ & $0.060$ & $0.090$ & {$1.000$} & {$1.000$}  & $0.982$ & {$0.987$} \\
& $H_{5,0.50}$    &$0.258$ & $0.043$ & $0.108$ & {$1.000$} & {$1.000$} & {$1.000$} & {$1.000$} \\
& $H_{6,0.50}$    &{$1.000$} & {$1.000$} & $0.709$ & {$1.000$} & {$1.000$} & {$1.000$} & {$1.000$} \\\midrule
$1000$
& $H_{1,0.00}$    &$0.059$ & {$0.053$} & $0.060$ & $0.056$ & $0.057$ & $0.061$ & $0.060$ \\
& $H_{2,0.00}$    &$0.043$ & $0.042$ & $0.070$ & $0.045$ & $0.046$  & $0.058$ & {$0.051$} \\
& $H_{3,0.00}$    &$0.063$ & {$0.054$} & $0.062$ & $0.057$ & {$0.054$} & {$0.038$} & $0.037$ \\
& $H_{4,0.00}$    &$0.059$ & {$0.053$} & $0.060$ & $0.056$ & $0.054$  & $0.061$ & $0.054$ \\
& $H_{5,0.00}$    &$0.055$ & $0.060$ & $0.047$ & $0.053$ & {$0.051$}  & $0.078$ & {$0.074$} \\
& $H_{6,0.00}$    &$0.045$ & {$0.047$} & $0.054$ & $0.057$ & $0.058$  & {$0.052$} & {$0.048$} \\\midrule
& $H_{1,0.25}$    &{$0.997$} & $0.992$ & $0.084$ & $0.938$ & $0.947$ & $0.730$ & {$0.747$} \\
& $H_{2,0.25}$    &$0.046$ & $0.067$ & $0.109$ & {$1.000$} & $0.910$ & {$0.947$} & $0.123$  \\
& $H_{3,0.25}$    &{$1.000$} & {$1.000$} & $0.459$ & {$1.000$} & {$1.000$}  & {$1.000$} & {$1.000$}  \\
& $H_{4,0.25}$    &$0.061$ & $0.052$ & $0.074$ & {$1.000$} & {$1.000$}  & $0.989$ & {$0.991$}  \\
& $H_{5,0.25}$    &$0.129$ & $0.059$ & $0.059$ & $0.993$ & $0.995$  & $0.936$ & {$0.971$}  \\
& $H_{6,0.25}$    &{$1.000$} & {$1.000$} & $0.257$ & {$1.000$} & {$1.000$} & {$1.000$} & {$1.000$}  \\\midrule
& $H_{1,0.50}$    &{$1.000$} & {$1.000$} & $0.281$ & {$1.000$} & {$1.000$} & {$1.000$} & {$1.000$}  \\
& $H_{2,0.50}$    &$0.049$ & $0.125$ & $0.305$ & {$1.000$} & {$1.000$} & {$1.000$} & {$1.000$} \\
& $H_{3,0.50}$    &{$1.000$} & {$1.000$} & $0.954$ & {$1.000$} & {$1.000$} & {$1.000$} & {$1.000$}  \\
& $H_{4,0.50}$    &$0.058$ & $0.057$ & $0.099$ & {$1.000$} & {$1.000$} & {$1.000$} & {$1.000$}  \\
& $H_{5,0.50}$    &$0.486$ & $0.058$ & $0.106$ & {$1.000$} & {$1.000$} & {$1.000$} & {$1.000$}  \\
& $H_{6,0.50}$    &{$1.000$} & {$1.000$} & $0.751$ & {$1.000$} & {$1.000$} & {$1.000$} & {$1.000$}   \\
\bottomrule\bottomrule
\end{tabular}
\caption{\small Proportion of rejections for the $R^2_n$, $U_n$, $\lambda^*_{4n}$, $T_n^{\mathrm{LCV}}$ and $T_n^{\mathrm{BLCV}}$ tests of independence for sample sizes $n=500,1000$ for a nominal significance level of $5\%$. For the six different models the values of the deviation from independence parameter are $\delta=0$ (independence), $0.25$ and $0.50$. Each proportion was calculated using $B=1000$ permutations for each of $M=1000$ random samples of size $n$ simulated from the specified model.  \label{tab1b}}
\end{table}

\begin{table}[htpb!]
\centering
\small
\begin{tabular}{cc|rrrrr}\toprule\toprule
\multirow{2}{*}{Test} & \multirow{2}{*}{$q$} & \multicolumn{5}{c}{Sample size} \\\cmidrule(lr){3-7}
& & $50$ & $100$ & $200$ & $500$ & $1000$ \\\midrule
$T_n^\mathrm{LCV}$ 	& $1$ & $0.17$ & $0.25$ & $0.93$ & $6.15$ & $28.41$ \\
 					& $2$ & $0.09$ & $0.27$ & $0.93$ & $6.14$ & $28.83$ \\\midrule
$T_n^\mathrm{BLCV}$ & $1$ & $0.66$ & $1.11$ & $2.05$ & $9.12$ & $33.79$ \\
					& $2$ & $4.27$ & $5.18$ & $9.98$  & $26.27$ & $71.89$ \\\bottomrule\bottomrule
\end{tabular}
\caption{\small Computing times (in seconds) for $T_n^\mathrm{LCV}$ and $T_n^\mathrm{BLCV}$ as a function of sample size and dimension $q$, with $q=1$ for the circular-linear case and $q=2$ for the spherical-linear case. The tests were run with $B=1000$ permutations and the times were measured in a $3.5$ GHz core.\label{tab3}}
\end{table}

Regarding power, the test for $R_n^2$ is the most powerful one for M1 and M3, although the performance of $T_n^\mathrm{LCV}$ and $T_n^\mathrm{BLCV}$, specially for M3, is quite similar. This was to be expected, as the circular-linear association tests should present more power against deviations of the first order. However, for M2, M4 and M5, all these tests are not able to distinguish the alternatives and the rejection ratios are close to the nominal level, resulting in $\lambda_{4n}^*$ being the test with better behavior among them. In contrast, $T_n^\mathrm{LCV}$ and $T_n^\mathrm{BLCV}$ correctly detect the deviations from the null. In M6, $R_n^2$ is only the most competitive for the situation with $n=50$, with $T_n^\mathrm{LCV}$ and $T_n^\mathrm{BLCV}$ the most competitive for the remainder of situations. $U_n$ shows a similar performance to $R_n^2$, but with more power in M2 and less in M5. $\lambda_{4n}^*$ is less affected than $R_n^2$ and $U_n$ by the change of models, but also has lower power than them for M1, M3 and M6. The results for the spherical-linear case are quite similar to the previous ones for the empirical size, but with lower power in comparison with the circular-linear scenario, something expected as a consequence of the difference in dimensionality.\\

Some final comments on the simulation results follow. For the different sample sizes and dimensions, the running times for $T_n^\mathrm{LCV}$ and $T_n^\mathrm{BLCV}$ are collected in Table \ref{tab3}. Computation times for $T_n^\mathrm{LCV}$ are very similar for different dimensions $q$, whereas $T_n^\mathrm{BLCV}$ is affected by $q$ due to the choice of the bandwidths $(h,g)_\mathrm{BLCV}$. The choice of the kernels was corroborated to be non-important for testing, as similar results were obtained for the test $T_n^\mathrm{LCV}$ using the directional-linear kernel $LK(r,t)=(1-r)\mathbbm{1}_{[0,1]}(r)\times \frac{3}{4}(1-t^2)\mathbbm{1}_{[-1,1]}(t)$. Cross-validatory bandwidths LSCV and BLSCV were also tried in the simulation study, providing worse results (this is also what usually happens with directional data, as it can be seen in \cite{Garcia-Portugues:exact}). Finally, it is worth mentioning that bootstrap calibration was also tried as an alternative to the permutation approach, using a pair of bandwidths for estimation and another pair for the smooth resampling. The results in terms of size, power and computing times were substantially worse than the ones obtained for permutations.\\

In conclusion, both $T_n^\mathrm{LCV}$ and $T_n^\mathrm{BLCV}$ tests show a competitive behavior in all the simulation models, sample sizes and dimensions considered, only being outperformed by $R_n^2$ in M1 and M3. Nevertheless, for those models, the rejection rates of both tests are in general close to the ones of $R_n^2$. The test $T_n^\mathrm{BLCV}$ corrects the over rejection of $T_n^\mathrm{LCV}$ in certain simulation models, without a significant loss in power but at the expense of a high computational cost. Finally, the classical tests $R_n^2$, $U_n$ and $\lambda_{4n}^*$ presented critical problems on detecting second order and some first order deviations from the independence. For all those reasons, the final recommendation is to preferably use the test $T_n^\mathrm{BLCV}$ for inference on directional-linear independence and $T_n^\mathrm{LCV}$ for a\nopagebreak[4] less computing intensive exploratory analysis.

\section{Real data analysis}
\label{sec:realdata}

\subsection{Data description}
\label{subsec:datadescription}

The original Portuguese fire atlas, covering the period from 1975 to 2005, is the longest annual and country-wide cartographic fire database in Europe \citep{Pereira}. Annual wildfire maps were derived from Landsat data, which represents the world's longest and continuously acquired collection of moderate resolution land remote sensing data, providing a unique resource for those who work in forestry, mapping and global change research. For each year in the dataset, Landsat imagery covering Portugal's mainland was acquired after the end of the fire season, thus providing a snapshot of the fires that occurred during the season. Annual fire perimeters were derived through a semi-automatic procedure that starts with supervised image classification, followed by manual editing of classification results. Minimum Mapping Unit (MMU), \textit{i.e.}, the size of the smallest fire mapped, changed according to available data. Between 1975 and 1983 (the MultiSpectral Scanner era), spatial resolution of satellite images is $80$ meters and MMU of $35$ hectares. From 1984 onwards with data availability at spatial resolution of $30$ meters (Thematic Mapper and Enhanced Thematic Mapper era) MMU is $5$ hectares, allowing to map a larger number of smaller fires than in the 1975--1983 era. Below an MMU of approximately $5$ hectares the burnt area classification errors increase substantially, and given the very skewed nature of fire size distribution, the $5$ hectares threshold ensures that over $90\%$ of total area actually burnt is mapped. For consistency, and due to discrepancies in minimum mapping unit between 1975--1983 and 1985--2005, in this study only fire perimeters mapped in the latter period were considered, which results in $26870$ fire perimeters.\\

This application is based on the watershed delineation proposed by \cite{Barros:2012p1135}. In their work, watersheds were derived from the Shuttle Radar Topography Mission (SRTM) digital terrain model \citep{Farr} using the ArcGIS hydrology toolbox \citep{ESRI}. Minimum watershed size was interactively increased so that each watershed contained a minimum of $25$ fire observations (see the cited work for more details). Fire perimeters straddling watershed boundaries were allocated to the watershed that contained its centroid.  \\

The orientation of fire perimeters and watersheds was determined by principal component analysis, following the approach proposed by \citet[pages 131--136]{Luo}. Specifically, principal component analysis was applied to the points that constitute the object's boundary (fire or watershed), with orientation given by the first principal component (PC1). Boundary points can be represented either in bidimensional space defined by each vertex's latitude and longitude coordinates, or in tridimensional space, taking also into account the altitude. Then, the PC1 corresponds to an axis that passes through the center of mass of the object and maximizes the variance of the projected vertices, represented in $\R^2$ or in $\R^3$. The fact of computing the PC1 also in $\R^3$ aims to take into account the variability of fires according to their slope, which, as the center plot of Figure \ref{fig3} shows, presents marked differences between regions. Then, the orientation of the object is taken as the direction given by its PC1. \\

It is important to notice that an orientation is an axial observation, and that some conversion is needed for applying the directional-linear independence test. In the two-dimensional case, the orientations can be encoded by an angular variable $\Theta\in[0,\pi)$, with period $\pi$, so $2\Theta$ is a circular variable. Then, with this codification, the angles $0$, $\frac{\pi}{2}$, $\pi$, $\frac{3\pi}{2}$ represent the E/W, NE/SW, N/S and NW/SE orientations, respectively. In the three-dimensional space, the orientation is coded by a pair of angles $(\Theta,\Phi)$ using spherical coordinates, where $\Theta\in[0,\pi)$ plays the same role as the previous setting and $\Phi\in[0,\frac{\pi}{2}]$ measures the inclination ($\Phi=\frac{\pi}{2}$ for flat slope and $\Phi=0$ for vertical; only positive angles are considered as the slope of a certain angle $\omega$ equals the slope of $-\omega$). Therefore, points with spherical coordinates $(2\Theta,\Phi)$, which lie on the upper semisphere, can be regarded as a realization of a spherical variable.

\subsection{Results}
\label{subsec:results}

The null hypothesis of independence between wildfire orientation and its burnt area (in log scale) is rejected, either using orientations in $\R^2$ or in $\R^3$, with a common $p$-value $0.000$. The test is carried out using the bandwidth selector BLCV (considered from now on) and all the $26870$ observations for years 1985--2005, ignoring stratification by watershed, and with $B=1000$ permutations. The $p$-values for the null hypothesis of independence between the orientation of a watershed and the total burnt area of fires within the region are $0.008$ and $0.000$ for orientations in $\R^2$ and in $\R^3$, respectively. Therefore, the null hypothesis is emphatically rejected. \\

\begin{figure}[htb]
\centering
\includegraphics[width=0.24\textwidth]{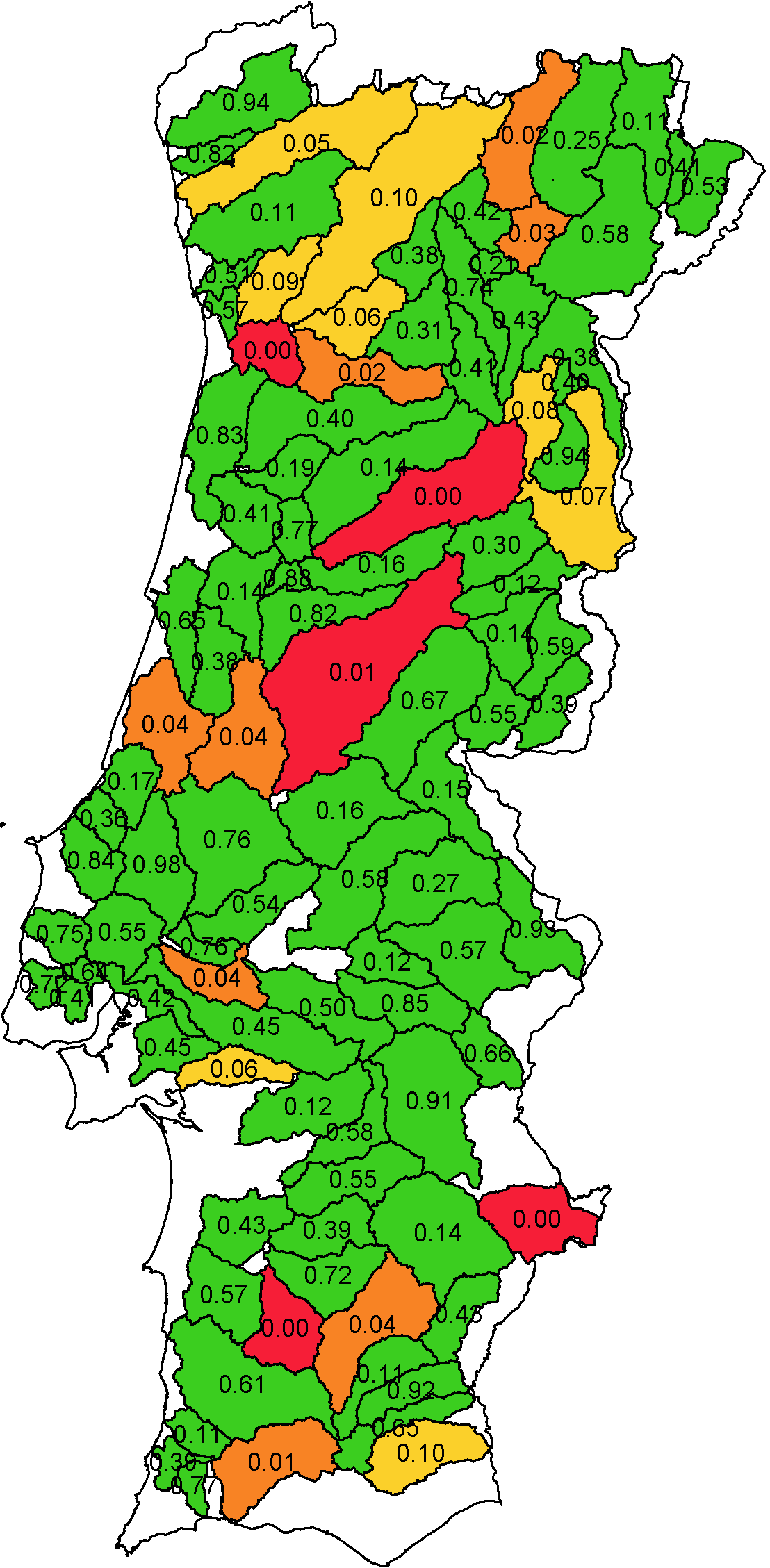}\includegraphics[width=0.24\textwidth]{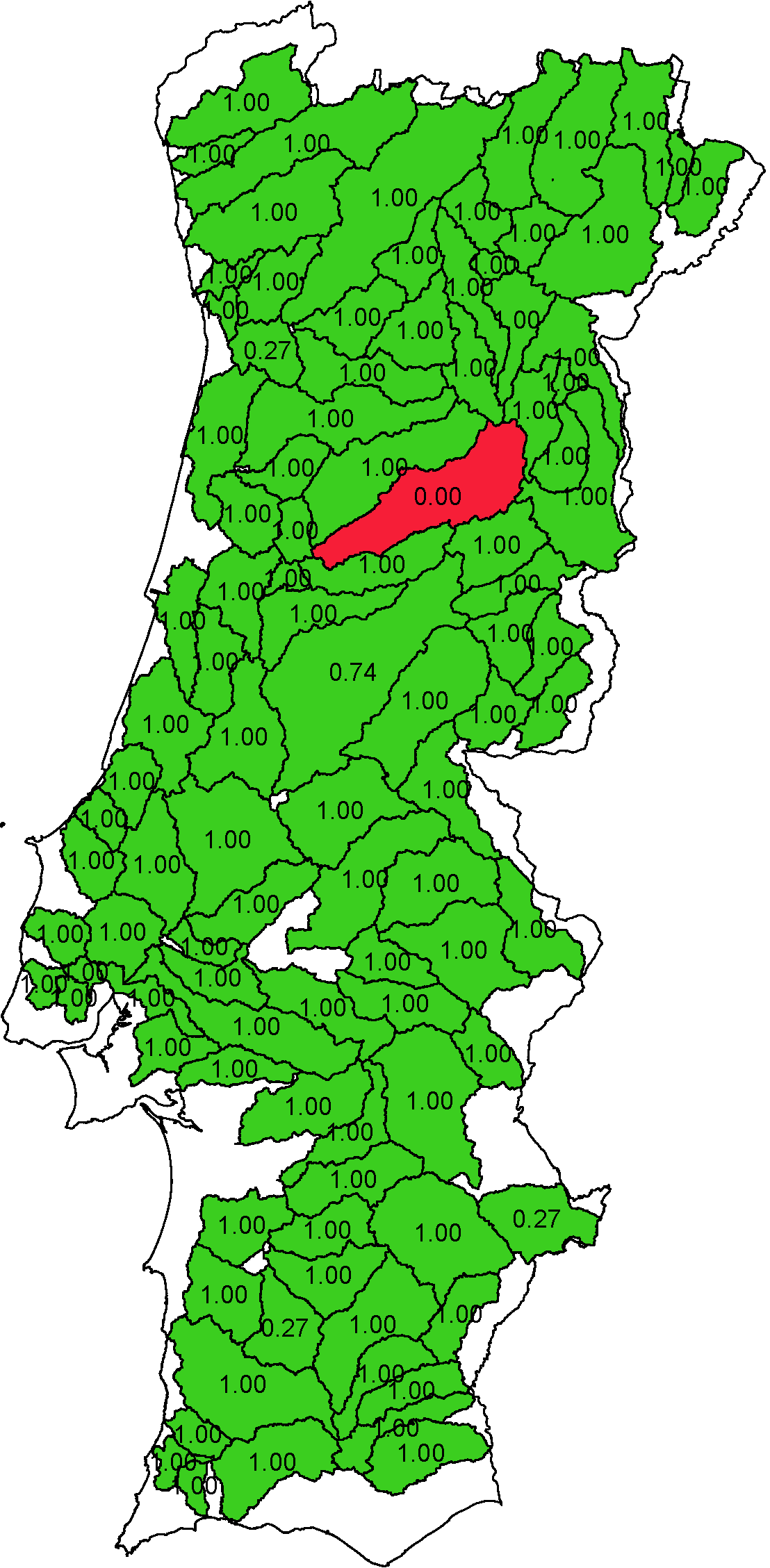}\includegraphics[width=0.24\textwidth]{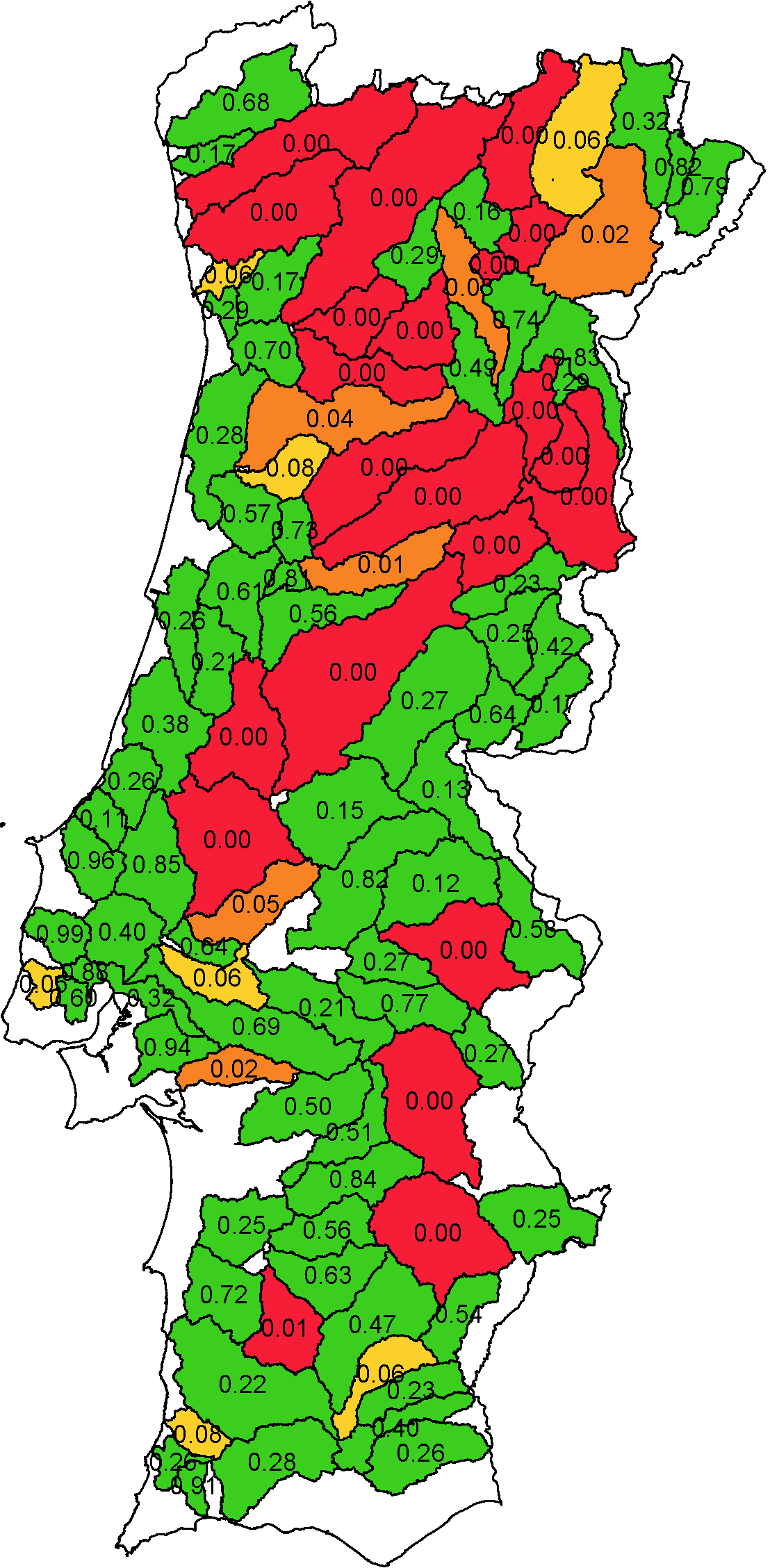}\includegraphics[width=0.24\textwidth]{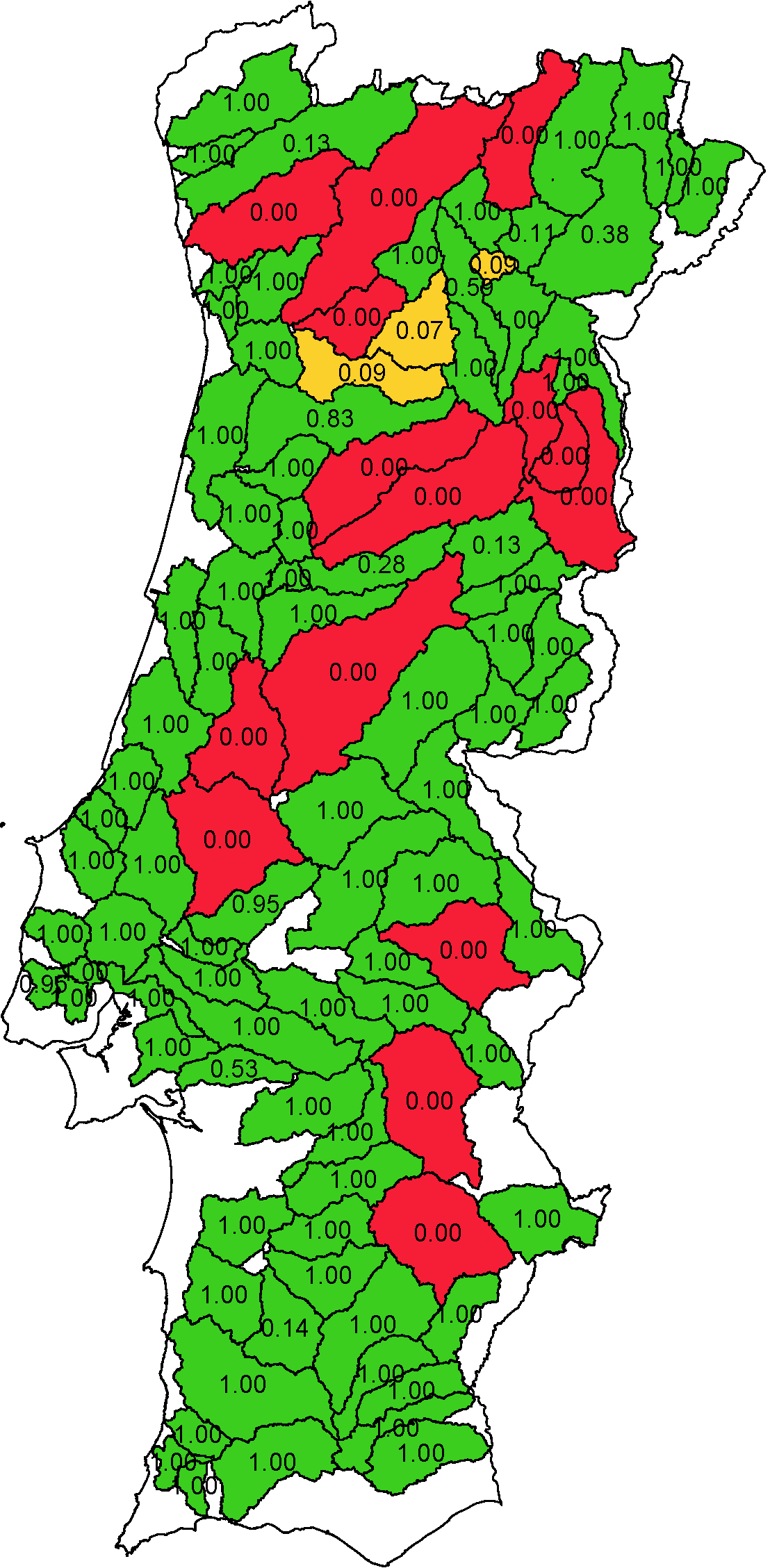}\\[0.3cm]
\includegraphics[scale=0.5]{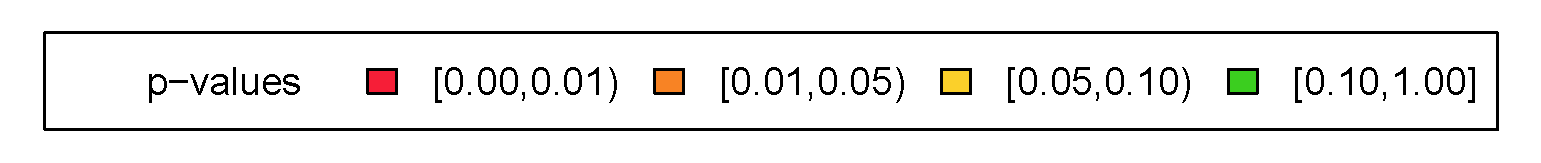}
\caption{\small $p$-values from the independence test for the first principal component PC1 of the fire perimeter and the burnt area (on a log scale), by watersheds. From left to right, the first and second maps represent the circular-linear $p$-values (PC1 in $\R^2$) and their corrected versions using the FDR, respectively. The third and fourth maps represent the spherical-linear situation (PC1 in $\R^3$), with uncorrected and corrected $p$-values by FDR, respectively. }
\label{fig4}       %
\end{figure}

\begin{figure}[h]
	\centering
	\includegraphics[width=0.49\textwidth]{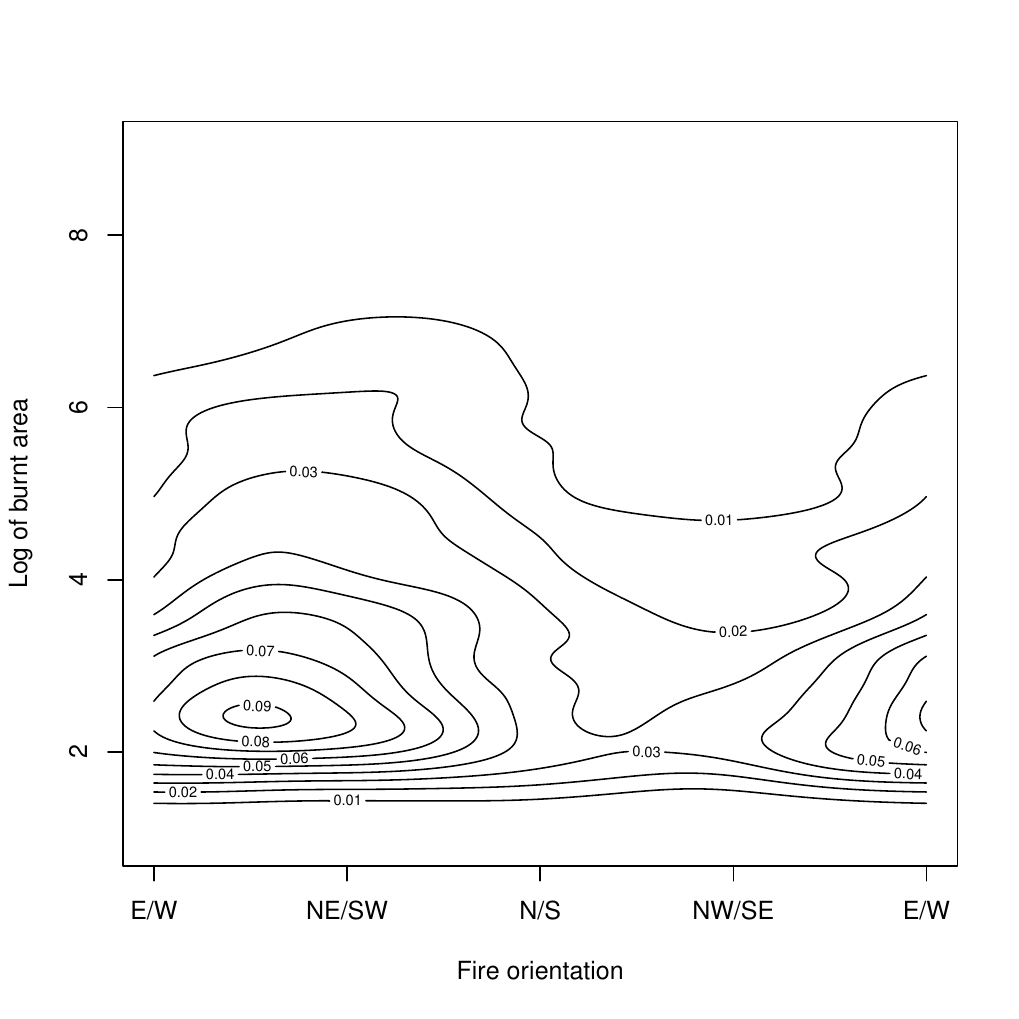}
	\includegraphics[width=0.49\textwidth]{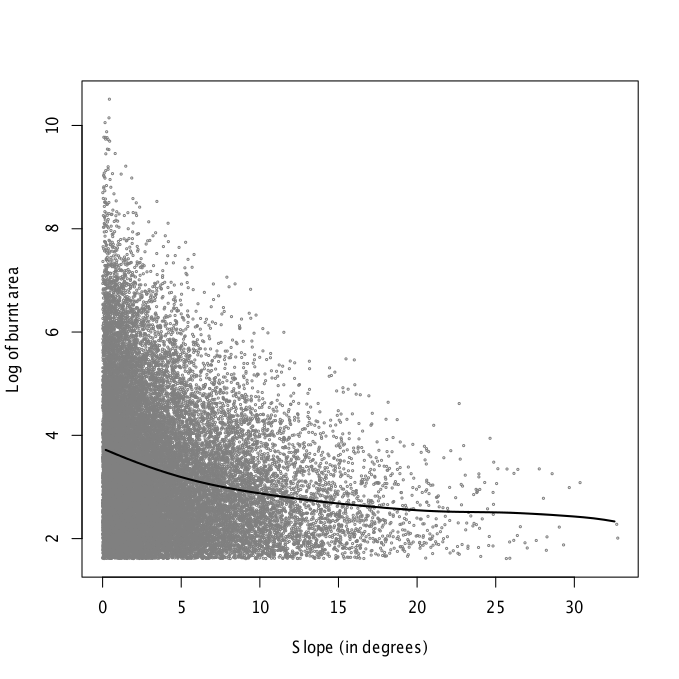}
	\caption{\small Left: density contour plot for fires in watershed number $31$, the watershed in the second plot on the left of Figure \ref{fig4} with $p\text{-value}=0.000$. The number of fires in the watershed is $n=1543$. The contour plot shows that the size of the area burnt is related with the orientation of the fires in the watershed. Right: scatter plot of the fires slope and the burnt area for the whole dataset, with a nonparametric kernel regression curve showing the negative correlation between fire slope and size. }
	\label{fig5}       %
\end{figure}

After identifying the presence of dependence between wildfire orientation and size, it is possible to carry out a watershed-based spatial analysis by applying the test to each watershed, in order to detect if the presence of dependence is homogeneous, or if it is only related to some particular areas. Figure \ref{fig4} represents maps of $p$-values of the test applied to the observations of each watershed, using PC1 in $\R^2$ and in $\R^3$ (from left to right, first and third plots of Figure \ref{fig4}, respectively). The maps reveal the presence of $13$ and $27$ watersheds where the null hypothesis of independence is rejected with significance level $\alpha=0.05$, for the circular-linear and the spherical-linear cases, respectively. This shows that the presence of dependence between fire orientation and size is not homogeneous and it is located in specific watersheds (see Figure \ref{fig5}). It is also interesting to note that the inclusion of the altitude coordinate in the computation of the PC1 leads to a richer detection of dependence between the wildfire orientation and size at the watershed level. This is due to the negative relation between the fire slope and size (see Figure \ref{fig5}), as large fires tend to have a flatter PC1 in $\R^3$ because they occur over highly variable terrain. Finally, the resulting $p$-values from the watershed analysis can also be adjusted using the False Discovery Rate (FDR) procedure of \cite{Benjamini2001} (from left to right, second and fourth plots of Figure \ref{fig4}). It is also possible to combine the $p$-values of the unadjusted maps with the FDR to test for independence between the wildfire orientation and the log-burnt area. The resulting $p$-values are $0.000$ for the circular-linear and spherical-linear\nolinebreak[4] cases. 

\section{Discussion}
\label{sec:discussion}

A nonparametric test for assessing independence between a directional and a linear component has been proposed, and its finite sample performance has been investigated in a simulation study. Simulation results support a satisfactory behavior of the permutation test implemented with LCV and BLCV bandwidths, in most cases outperforming the available circular-linear testing proposals, and being competitive in other cases. The proposed BLCV bandwidths presents better results in terms of empirical size, although further study is required in bandwidth selection. In addition, when the null hypothesis of independence is rejected, the kernel density estimate can be used to explore the form of dependence, at least for the circular-linear and\nopagebreak[4] spherical-linear cases.\\

The application of the test to the entire wildfire orientation and size dataset makes possible the detection of dependence between these two variables, for both two-dimensional or three-dimensional orientation. The same conclusion holds for watershed orientation and total area burnt. A detailed study of each watershed allows for a more specific insight into the problem. The evidence of independence between fire size and fire orientation in some watersheds suggests that an event-based analysis (such as the work of \citet{Barros:2012p1135}) should yield results similar to those that would be expected from an area-based analysis. On the other hand, detection of dependence between fire size and orientation in watersheds with uniform orientation \citep{Barros:2012p1135} highlights cases where there may be a mixture of orientations. In such cases, an analysis taking fire size into account might find evidence of preferential orientation in fire perimeters. In watersheds where fire events show preferential orientation (non-uniform distribution) and there is dependence between size and orientation, fire orientation distributions are structured in relation to fire size, especially considering the typically asymmetric nature of fire size distributions, dominated by a small number of very large events \citep{strauss}. In these cases, an area-weighted analysis of fire perimeter orientation might lead to different results than those found by \citet{Barros:2012p1135}. When altitude is included in calculation of the PC1 in $\R^3$ it highlights the negative relation between fire slope and size,  which is mostly due to the fact that larger fires present flatter PC1. Slope has a skewed distribution, with low mean value and a relatively long right tail. Thus, while small fires usually occur on high slopes, large fires on consistently steep areas are unlikely.\\

Finally, it can be argued that the data are probably not independent and identically distributed over space and time. Unfortunately, given the data gathering procedure (detailed at the beginning of Section \ref{sec:realdata}) dependence patterns cannot be clearly identified. Accounting for temporal or spatial dependence directly in the directional-linear kernel estimator and in the testing procedure is an open\nolinebreak[4] problem.


\section*{Acknowledgments} 
The authors acknowledge the support of Project MTM2008--03010, from the Spanish Ministry of Science and Innovation, Project 10MDS207015PR from Direcci\'on Xeral de I+D, Xunta de Galicia and IAP network StUDyS, from Belgian Science Policy. Work of E. Garc\'ia-Portugu\'es has been supported by FPU grant AP2010--0957 from the Spanish Ministry of Education and work of A. M. G. Barros, by Ph.D. Grant SFRH/BD/40398/2007 from the Fun\-da\c{c}\~{a}o para a Ci\^{e}ncia e Tecnologia. J. M. C. Pereira participated in this research under the framework of research projects ``Forest fire under climate, social and economic changes in Europe, the Mediterranean and other fire-affected areas of the world (FUME)'', EC FP7 Grant Agreement No. 243888 and ``Fire-Land-Atmosphere Inter-Relationships: understanding processes to predict wildfire regimes in Portugal'' (FLAIR), PTDC/AAC/AMB/104702/2008. Authors gratefully acknowledge an anonymous referee for the suggestion of employing permutations for the test calibration and the careful revision of the paper. Authors also acknowledge the\nopagebreak[4] suggestions raised by another referee.

\appendix

\section{Proof of Lemma \ref{lem:1}}
\label{sec:appendixA}

\begin{proof}
The closed expression (just involving matrix computations) for $T_n$ is obtained by splitting the calculus into three addends:
\begin{align*}
T_n=&\,\Iqr{\lrp{\hat f_{(\bX,Z);h,g}(\bx,z)-\hat f_{\bX;h}(\bx)\hat f_{Z;g}(z)}^2}{\bx}{z}\\
=&\,\Iqr{\bigg(\frac{c_{h,q}(L)}{ng}\sum_{i=1}^n L\lrp{\frac{1-\bx^T\bX_i}{h^2}}K\lrp{\frac{z-Z_i}{g}}-\hat f_{\bX;h}(\bx)\hat f_{Z;g}(z)\bigg)^2}{\bx}{z}\\
=&\,\sum_{i=1}^n\sum_{j=1}^n\Iqr{\frac{c_{h,q}(L)^2}{n^2g^2} L\lrp{\frac{1-\bx^T\bX_i}{h^2}}K\lrp{\frac{z-Z_i}{g}} L\lrp{\frac{1-\bx^T\bX_j}{h^2}}K\lrp{\frac{z-Z_j}{g}}}{\bx}{z}\\
&-2\sum_{i=1}^n\Iqr{\frac{c_{h,q}(L)}{ng} L\lrp{\frac{1-\bx^T\bX_i}{h^2}}K\lrp{\frac{z-Z_i}{g}}\hat f_{\bX;h}(\bx)\hat f_{Z;g}(z)}{\bx}{z}\\
&+\Iqr{\hat f_{\bX;h}(\bx)^2\hat f_{Z;g}(z)^2}{\bx}{z}\\
=&\,(\co)-(\co)+(\co).
\end{align*}
The first addend is\addtocounter{equation}{-2}
\begin{align*}
(\arabic{equation})=&\,\frac{c_{h,q}(L)^2}{n^2g^2}\sum_{i=1}^n\sum_{j=1}^n\Iqr{ L\lrp{\frac{1-\bx^T\bX_i}{h^2}}K\lrp{\frac{z-Z_i}{g}} L\lrp{\frac{1-\bx^T\bX_j}{h^2}}K\lrp{\frac{z-Z_j}{g}}}{\bx}{z}\\
=&\,\frac{c_{h,q}(L)^2}{n^2g^2}\sum_{i=1}^n\sum_{j=1}^n\Iq{e^{-2/h^2}e^{\bx^T(\bX_i+\bX_j)/h^2}}{\bx}\times\Ir{K\lrp{\frac{z-Z_i}{g}}K\lrp{\frac{z-Z_j}{g}}}{z}\\
=&\,\frac{c_{h,q}(L)^2}{n^2}e^{-2/h^2}\sum_{i=1}^n\sum_{j=1}^n\frac{\phi_{\sqrt{2}g}(Z_i-Z_j)}{C_q\lrp{\norm{\bX_i+\bX_j}/h^2}}\\
=&\,\frac{C_q\lrp{1/h^2}^2}{n^2}\sum_{i=1}^n\sum_{j=1}^n\frac{\phi_{\sqrt{2}g}(Z_i-Z_j)}{C_q\lrp{\norm{\bX_i+\bX_j}/h^2}}.
\end{align*}
For the second addend,\addtocounter{equation}{+1}
\begin{align*}
(\arabic{equation})=&\,2\sum_{i=1}^n\Iqr{\frac{c_{h,q}(L)}{ng} L\lrp{\frac{1-\bx^T\bX_i}{h^2}}K\lrp{\frac{z-Z_i}{g}}\hat f_{\bX;h}(\bx)\hat f_{Z;g}(z)}{\bx}{z}\\
=&\,2\frac{c_{h,q}(L)}{ng}\sum_{i=1}^n\Iq{L\lrp{\frac{1-\bx^T\bX_i}{h^2}}\hat f_{\bX;h}(\bx)}{\bx}\times\Ir{K\lrp{\frac{z-Z_i}{g}}\hat f_{Z;g}(z)}{z}\\
=&\,2\frac{c_{h,q}(L)}{ng}\sum_{i=1}^n\Bigg\{\Bigg[\Iq{L\lrp{\frac{1-\bx^T\bX_i}{h^2}}\frac{c_{h,q}(L)}{n}\sum_{j=1}^n L\lrp{\frac{1-\bx^T\bX_j}{h^2}}}{\bx}\Bigg]\\
&\times\lrc{\Ir{K\lrp{\frac{z-Z_i}{g}}\frac{1}{ng}\sum_{k=1}^n K\lrp{\frac{z-Z_k}{g}}}{z}}\Bigg\}\\
=&\,\frac{2}{n^3}\sum_{i=1}^n\Bigg\{\Bigg[\Iq{c_{h,q}(L)^2e^{-2/h^2} e^{\bx^T\lrp{\bX_i+\bX_j}/h^2}}{\bx}\Bigg]\times\lrc{\sum_{k=1}^n\phi_{\sqrt{2}g}(Z_i-Z_k)}\Bigg\}\\
=&\,\frac{2}{n^3}\sum_{i=1}^n\Bigg\{\Bigg[\sum_{j=1}^n \frac{C_q\lrp{1/h^2}^2}{C_q\lrp{\norm{\bX_i+\bX_j}/h^2}}\Bigg]\times\lrc{\sum_{k=1}^n\phi_{\sqrt{2}g}(Z_i-Z_k)}\Bigg\}.
\end{align*}
Finally, the third addend is obtained as\addtocounter{equation}{+1}
\begin{align*}
(\arabic{equation})=&\,\Iqr{\hat f_{\bX;h}(\bx)^2\hat f_{Z;g}(z)^2}{\bx}{z}\\
=&\,\Iq{\hat f_{\bX;h}(\bx)^2}{\bx}\times\Ir{\hat f_{Z;g}(z)^2}{z}\\
=&\,\Bigg[\sum_{i=1}^n\sum_{j=1}^n\frac{c_{h,q}(L)^2}{n^2}\Iq{L\lrp{\frac{1-\bx^T\bX_i}{h^2}} L\lrp{\frac{1-\bx^T\bX_j}{h^2}}}{\bx}\Bigg]\\
&\times\Bigg[\sum_{i=1}^n\sum_{j=1}^n\frac{1}{n^2g^2}\Ir{ K\lrp{\frac{z-Z_i}{g}}K\lrp{\frac{z-Z_j}{g}}}{z}\Bigg]\\
=&\,\Bigg[\frac{1}{n^2}\sum_{i=1}^n\sum_{j=1}^n\frac{C_q\lrp{1/h^2}^2}{C_q\lrp{\norm{\bX_i+\bX_j}/h^2}}\Bigg]\times \Bigg[\frac{1}{n^2}\sum_{i=1}^n\sum_{j=1}^n\phi_{\sqrt{2}g}(Z_i-Z_j)\Bigg].
 \end{align*}
From the previous results and after applying some matrix algebra, it turns out that
\begin{align*}
T_n=&\,\mathbf{1}_n\bigg(\frac{1}{n^2}\mathbf{\Psi}(h)\circ\mathbf{\Omega}(g)-\frac{2}{n^3}\mathbf{\Psi}(h)\mathbf{\Omega}(g)+\frac{1}{n^4}\mathbf{\Psi}(h)\mathbf{1}_n\mathbf{1}_n^T\mathbf{\Omega}(g)\bigg)\mathbf{1}_n^T,
\end{align*}
where:
\begin{align*}
\mathbf{\Psi}(h)=\lrp{\frac{C_q\lrp{1/h^2}^2}{C_q\lrp{\norm{\bX_i+\bX_j}/h^2}}}_{ij},\quad\mathbf{\Omega}(g)=\lrp{\phi_{\sqrt{2}g}\lrp{Z_i-Z_j}}_{ij}.
\end{align*}
\end{proof}

\renewcommand{\bibname}{References}

\end{document}